\documentclass[aps,prb,amsmath,amssymb,twocolumn,floatfix,showpacs,superscriptaddress]{revtex4-1}       

\usepackage{graphicx,bm}                                           
\usepackage[usenames,dvipsnames]{color}                    
\usepackage[colorlinks=true]{hyperref}                          
\hypersetup{    
    bookmarks=true,         
    unicode=false,          
    pdftoolbar=true,        
    pdfmenubar=true,        
    pdffitwindow=false,     
    pdfstartview={FitH},    
    pdftitle={My title},    
    pdfauthor={Author},     
    pdfsubject={Subject},   
    pdfcreator={Creator},   
    pdfproducer={Producer}, 
    pdfkeywords={keyword1} {key2} {key3}, 
    pdfnewwindow=true,      
    colorlinks=true,       
    linkcolor=magenta, 
    citecolor=blue,        
    filecolor=magenta,      
    urlcolor=cyan           
} 
\usepackage{epstopdf}      
\usepackage{times,soul}          
\usepackage[tight]{subfigure}            
\usepackage{multirow,array}

\DeclareMathOperator{\tr}{Tr}

\renewcommand{\vec}[1]{\mathbf{#1}}

\newcommand{\Renyi}{R\'enyi\ }
\newcommand{\ren}{R\'enyi\ }
\newcommand{\amin}{\frak a_{\alpha}^{(M)}}
\newcommand{\req}[1]{Eq.\thinspace(\ref{#1})}

\newcommand\Ts{\rule{0pt}{2.8ex}}       
\newcommand\Bs{\rule[-1.3ex]{0pt}{0pt}} 

\begin{document}

\title{Universal corner entanglement of Dirac fermions and gapless bosons\\ from the continuum to the lattice} 
\author{Johannes Helmes}   
\affiliation{Institute for Theoretical Physics, University of Cologne, 50937 Cologne, Germany}
\affiliation{Perimeter Institute for Theoretical Physics, Waterloo, Ontario, N2L 2Y5, Canada}
\author{Lauren E. Hayward Sierens}
\affiliation{Perimeter Institute for Theoretical Physics, Waterloo, Ontario, N2L 2Y5, Canada}
\affiliation{Department of Physics and Astronomy, University of Waterloo, Ontario, N2L 3G1, Canada}

\author{Anushya Chandran}
\affiliation{Perimeter Institute for Theoretical Physics, Waterloo, Ontario, N2L 2Y5, Canada}
\author{William Witczak-Krempa}
\affiliation{Department of Physics, Harvard University, Cambridge, MA 02138, USA}
\author{Roger G. Melko}
\affiliation{Perimeter Institute for Theoretical Physics, Waterloo, Ontario, N2L 2Y5, Canada}
\affiliation{Department of Physics and Astronomy, University of Waterloo, Ontario, N2L 3G1, Canada}

\date{\today}

\begin{abstract}    

A quantum critical (QC) fluid exhibits universal subleading corrections to the area law of its entanglement entropies.
In two dimensions when the partition involves a corner of angle $\theta$, the subleading term is logarithmic with coefficient 
$a_\alpha(\theta)$ for the $\alpha$-R\'enyi entropy.  
In the smooth limit $\theta\!\to\!\pi$, $a_1(\theta)$ yields the central charge of   
the stress tensor when the QC point is described by a conformal field theory (CFT). For general  
\ren indices and angles, $a_\alpha(\theta)$ is richer and few general results exist. 
We study $a_\alpha(\theta)$ focusing on two benchmark CFTs, 
the free Dirac fermion and boson.  
We perform numerical lattice calculations to obtain high precision results in $\theta,\alpha$ regimes hitherto unexplored. 
We derive field theory estimates for $a_\alpha(\theta)$, including new exact results, 
and demonstrate an excellent quantitative match with our numerical
calculations. 
We also develop and test strong lower bounds, 
which apply to both free and interacting QC systems. Finally, we comment on the near collapse of $a_\alpha(\theta)$
for various theories, including interacting $O(N)$ models. 

\end{abstract}

\maketitle 
\tableofcontents        
 
%
%

\section{Introduction}

Quantum entanglement provides a valuable characterization of the universal low-energy physics of diverse quantum many-body systems \cite{Eisert:2010ab}.  
At quantum critical points (QCPs) in particular, entanglement measures contain universal functions that only depend on the various scale-invariants of the geometry of the entangling bipartition.
In $d=1$ spatial dimension for example, the entanglement entropy of a sub-region diverges logarithmically with the size of the sub-region, with a coefficient that measures the central charge of the 
associated conformal field theory (CFT).\cite{Holzhey:1994, Vidal:2003aa, Korepin:2004aa, Calabrese:2004}  
In higher dimensions, measures of entanglement are crucially sensitive to the non-trivial geometries of the sub-region.
Recent work \cite{Atheorem,Fradkin:2006,Metlitski:2009,Casini:2009,MyersSinha1,MyersSinha2,Max_Tarun,CHM_2011,Jafferis2011,Casini:2012,Grover:2014e,Myers:2012,Stephan_2013,MIFtheorem} has explored the universal parts of these entropies, and even identified quantities that play the role of the central charge in higher dimensions by organizing critical theories under the renormalization group flow.
The possibility of a quantum non-local property like entanglement acting to organize interacting critical theories in all dimensions is an exciting prospect. 

\begin{figure}[t]
    \centering
    \includegraphics[width=0.45\linewidth]{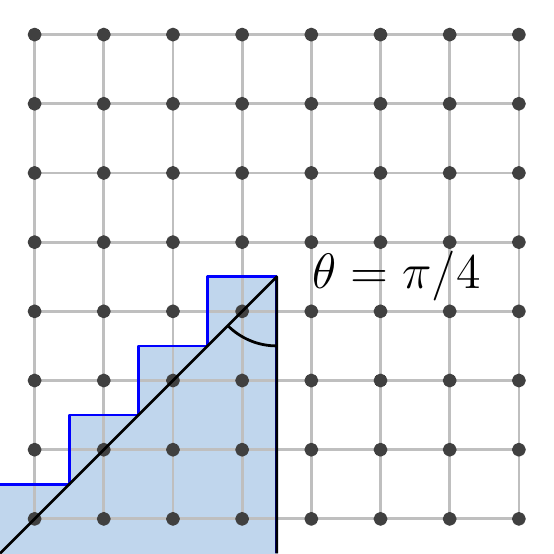}
    \hspace{0.5cm}\includegraphics[width=0.45\linewidth]{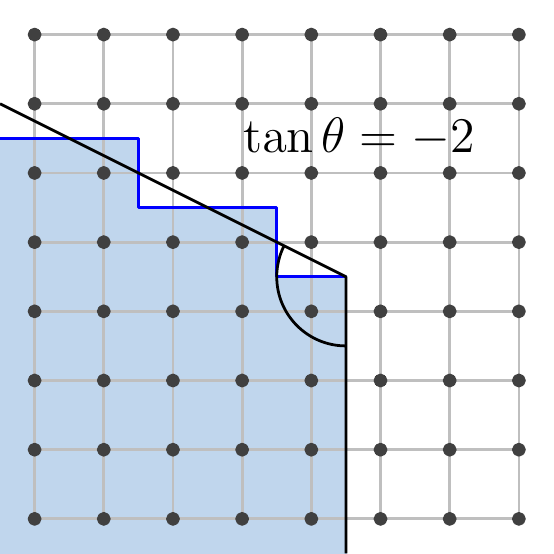}
    \caption{Examples of corners in the entangling boundary, superimposed on an underlying square lattice, 
    with opening angles of $\theta = \pi /4 $ (left panel) and $\tan \theta = -2$ (right panel).}
    \label{fig:corner45}
\end{figure}

In this paper, we study the entanglement of critical systems in two spatial dimensions ($d=2$).  
We focus on bipartitions that contain one or more corners, which arise naturally in lattice models and are easily accessible numerically.     
Consider a sub-region $A$ of linear dimension $\ell$ with a single corner with opening angle $\theta$ (Fig.~\ref{fig:corner45}).
A frequently-used measure of the entanglement between $A$ and its complement $\bar{A}$ is the \ren entropy with index $\alpha$, given by
\begin{align} 
S_\alpha(A) \equiv \frac{1}{1-\alpha} \ln(\textrm{Tr } \rho_A^\alpha),
\end{align}
where $\rho_A$ is the reduced density matrix of the sub-region $A$ in the ground state.
Note that $S_1(A)$ is the familiar von Neumann entanglement entropy.
In a gapless system without a Fermi surface, it is known 
that $S_\alpha(A)$ scales with the size of $A$ according to
\begin{equation} \label{eq:S_corner}
S_\alpha = \mathcal A_\alpha \frac{\ell}{\delta} - a_\alpha (\theta) \ln \Big( \frac{\ell}{\delta} \Big)+ \mathcal{O}(1)\,,
\end{equation}
where $\delta$ is a lattice length scale, $\mathcal A_\alpha$ is the non-universal ``area-law'' coefficient, 
and the sub-leading logarithmic term is a consequence of the corner in the boundary of $A$.

The corner coefficient $a_\alpha(\theta)$ is universal, 
being independent of the short distance cutoff $\delta$, and constitutes the focus of this article. 
Recent numerical work on the interacting quantum critical theories that arise in the 2$d$ Ising, 
XY and Heisenberg models\cite{Singh:2012t,Kallin:2013,Kallin:2014,Helmes:2014,Miles:2014,Helmes:2015} suggests that this corner coefficient provides a measure of the low energy degrees of freedom of the critical theory. 
Partially motivated by these developments, Refs.~\onlinecite{Bueno:2015,Bueno:2015b} conjectured that the central charge associated with the stress (energy-momentum) tensor       
of the CFT underlying the critical point is proportional to $a_1(\theta\!\to\!\pi)$. 
This relationship provides a way to estimate the central charge starting from a realistic (lattice) wavefunction.  
Subsequent works \cite{Elvang:2015,Miao2015,Faulkner:2015aa} proved this conjecture, providing an important relationship between the corner coefficient and a fundamental property of the CFT.   

Beyond the smooth angle limit, $\theta\!\to\! \pi$, little is known about the physical content of $a_\alpha(\theta)$ at general angles and \ren indices, even in free theories. 
Previous works \cite{Casini:2007,Casini:2008,Casini:2009,Sahoo} have computed $a_\alpha(\theta)$ in free theories at special angles like $\theta=\pi/2,\pi/4$ that are ``natural'' for a square lattice (see Fig.~\ref{fig:corner45})
In this article we explore the universal content of the corner coefficient for a variety of opening angles $\theta$ beyond those commensurate with the lattice in two prototypical free CFTs, the Dirac fermion and complex relativistic boson, using two different methods.
After first reviewing some general properties of the corner coefficient in Section \ref{sec:basics},
in Section \ref{sec:QFT} we derive a field theory approximation for $a_{\alpha}(\theta)$ by expanding around the smooth limit.  
From this, we obtain high-precision values for a variety of angles, as well as a rigorous new set of lower bounds, 
which apply to free and interacting theories. 
Then, in Section \ref{sec:latt} we solve the two theories exactly on finite lattices.    
Using detailed finite-size extrapolations, we obtain $a_\alpha(\theta)$ in the thermodynamic limit. 
We find excellent agreement between the lattice numerics and field theory results.
We also discover that $a_\alpha(\theta)$, when properly normalized, approximately collapses to a single
curve in a variety of theories. This occurs in a broad range of angles and \ren indices.
In addition, previous theoretical treatments \cite{Bueno:2015JHEP} pointed 
out a surprising duality between the corner coefficients of the complex boson and the Dirac 
fermion in the smooth angle limit.   
We find that this duality approximately holds far away from that limit, even up to $\theta=\pi/2$.
Taken together, our results indicate that calculations on natural lattice angles, such as $\theta = \pi/2$, may be enough to 
approximately study relationships derived in the smooth angle limit.
This bodes well for the continuing interplay between lattice numerics and quantum field theory in the geometrical study of entanglement in
correlated systems.

\section{Properties of the corner coefficient} \label{sec:basics}

We begin by reviewing some general properties of the corner coefficient.  
First, $a_\alpha(\theta)$ is symmetric about $\pi$ such that 
\begin{align}
\label{Eq:asymmetric}
a_\alpha(2\pi-\theta)=a_\alpha(\theta). 
\end{align}
This follows from the equality $S_\alpha(A) = S_{\alpha}(\bar{A})$ in any pure state. 
We therefore restrict our analysis to $0<\theta \leq \pi$ in this article. 
Second, it was shown\cite{Hirata:2007} that for CFTs, the corner function at $\alpha=1$ is decreasing and convex for $0<\theta\leq \pi$:
\begin{align} \label{basic}
  a_1'(\theta) \leq 0\, , \qquad a_1''(\theta)\geq 0\,.
\end{align}
Eq.~\eqref{basic} was derived using the strong sub-additivity (SSA) of the entanglement entropy $S_1(A)$ and Eq.~\eqref{Eq:asymmetric}. 
As the derivation does not assume Lorentz invariance, Eq.~(\ref{basic}) should hold at all scale invariant quantum critical points.
Indeed, they can be explicitly seen to hold for a special class of $z=2$ Lifshitz QCPs\cite{Fradkin:2006} (see also Ref.~\onlinecite{Bueno:2015JHEP}).  

Next, the corner function vanishes as $\theta\to \pi$ because the corner disappears in the smooth limit. In fact, we have   
\begin{align} \label{smooth1}
  a_\alpha(\theta\simeq\pi) = \sigma_\alpha\cdot (\theta-\pi)^2,
\end{align}
which follows because of 1) the reflection property Eq.~\eqref{Eq:asymmetric}, and 2) the expectation that the corner function is analytic about $\pi$. 
For general CFTs, the results of Ref.~\onlinecite{Bianchi:2015}
imply that \req{smooth1} holds for integer $\alpha\!\geq\! 2$.
Ref.~\onlinecite{Faulkner:2015aa} proved \req{smooth1} at $\alpha\!=\!1$, 
as well as the conjecture\cite{Bueno:2015} that the coefficient $\sigma_1$ is given by
\begin{align} \label{smooth-CT}    
  \sigma_1=\frac{\pi^2}{24}\, C_T\,.
\end{align}
$C_T$ is the central charge associated with the stress tensor $T_{\mu \nu}$ of the CFT:
$\langle T_{\mu \nu}(x) T_{\lambda \rho}(0) \rangle = C_T \mathcal{I}_{\mu \nu, \lambda \rho}(x) / |x|^{6}$, 
with $\mathcal{I}_{\mu \nu, \lambda \rho}$ being a dimensionless tensor structure fixed by conformal symmetry.\cite{Osborn:1994}  
When all the indices are set to the time-direction, the 2-point function describes the universal part of the
energy density correlations in the groundstate.  
For the complex boson and Dirac fermion,\cite{Osborn:1994} $C_T=3/(16\pi^2)$. 
 
Interestingly, when \req{smooth-CT} is combined with SSA, it leads to the general lower bound\cite{Bueno:2016} 
\begin{align} \label{ssa-bound}
  a_1(\theta)  \geq \frac{\pi^2 C_T}{3} \ln(1/\sin(\theta/2)) \,.
\end{align} 
For (unitary) CFTs, $C_T\geq 0$ implying in particular that the corner function is non-negative.
It was further found that even at $\theta\!=\! \pi/2$ (away from the smooth limit), $a_1(\pi/2)/C_T$  
takes approximately the same value in a wide range of CFTs.\cite{Bueno:2015,Bueno:2015b,Bueno:2016}  
Specifically, $a_1(\pi/2)/C_T \approx 1.2$, which nearly saturates the lower bound of $(\pi^2/6)\ln 2\!\approx\! 1.14$, \req{ssa-bound}.  
These observations help understand why $a_1(\pi/2)$ is seen to scale linearly with $N$ for the
$O(N)$ Wilson-Fisher QCPs\cite{Bueno:2015}: a conformal bootstrap calculation\cite{Kos:2014} has shown that $C_T^{O(N)}\!\simeq\!C_T^{O(1)} N$   
for $N=1,2,3$.\footnote{The relation is very accurate: $C_T^{O(2)}/(2C_T^{O(1)})=0.9969$ and $C_T^{O(3)}/(3C_T^{O(1)})=0.9974$.}  
This linear scaling of $C_T$ then translates to an approximate linear-in-$N$ scaling for $a_1(\pi/2)$.              

In the opposite limit of $\theta \to 0$, the lower bound \req{ssa-bound} implies that $a_1(\theta)$ diverges. 
The actual singularity is stronger than the logarithmic one found in \req{ssa-bound} and obeys\cite{Casini:2007}  
\begin{align}  \label{sharp}
  a_\alpha(\theta\simeq 0) = \kappa_\alpha / \theta\,, 
\end{align} 
where $\kappa_\alpha$ is a universal constant characterizing the system. 
To heuristically understand this divergence, 
let us imagine that the subregion $A$ is an isosceles triangle with an angle $\theta$ going to zero, as the 
adjoining segments, of fixed length, approach each other. 
The entropy of $A$ needs to vanish in that limit, 
 but the area law prefactor $\mathcal A_\alpha$, in \req{eq:S_corner}, remains finite. 
Since the corner contribution is negative, \req{eq:S_corner}, its growth as $\theta\!\to \!0$
will counteract the area law, and thus contribute to the vanishing of the entire entropy $S_\alpha(A)$.

For critical ground states described by CFTs, $\kappa_\alpha$ also dictates the universal part of the entanglement entropy of a thin strip as the width is sent to zero.\cite{Casini:2009,Bueno:2015b} 
As such, $\kappa_\alpha$ appears in the universal entanglement on torus and cylinder topologies,\cite{Will-torus,Fradkin_2015}
i.e.\/ in systems with periodic boundary conditions.    

\section{Field theory results}  \label{sec:QFT}
In this section, we use quantum field theory methods to obtain precise estimates and new lower bounds for the corner function. 
We test our methods using the the free relativistic boson and Dirac fermion.
The central idea is to expand the corner function about the smooth limit $\theta=\pi$ such that\cite{Casini:2009} 
\begin{align} \label{series}
  a_\alpha(\theta) = \sum_{p=1} \sigma_\alpha^{(p-1)}\cdot (\theta-\pi)^{2p}\,,
\end{align} 
where $\sigma_\alpha^{(p)}$ are the smooth limit expansion coefficients. (Note that we also use the 
notation $\sigma, \sigma',\sigma''$ for the first three coefficients.) 
Only even powers appear due to the reflection symmetry about $\pi$, \req{Eq:asymmetric}.
Below, we shall obtain most of the boson and fermion smooth limit coefficients by numerically evaluating 
integrals\cite{Casini:2007,Casini:2009}, although in certain cases we are able to derive new analytical answers. 
The $\sigma^{(p)}$ coefficients are not only easier to evaluate than the full $a(\theta)$, but also play a key role in 
obtaining lower bounds, as we now explain.

\subsection{General bounds} \label{sec:bounds} 
We obtain strong lower bounds for $a_\alpha(\theta)$ valid for general \ren index in a wide array of theories. 
The starting point is the reflection positivity property of Euclidean quantum field theory, which leads to   
an infinite set of non-linear differential inequalities for integer $\alpha\!\geq\! 2$ \ren entropies.\cite{Casini2012}
When applied to the corner function these can be simply expressed using a determinant of derivatives as\cite{Casini2012}
\begin{align} \label{RP}
  \det\, \big\{\partial_\theta^{j+k+2} a_\alpha(\theta) \big\}_{j,k=0}^{M} \, \geq 0\, ,  
\end{align}
where $M\geq 0$.
Expanding the determinants, the first two inequalities, $M=0 \text{ and } 1$, respectively read
\begin{align}
  \partial_\theta ^2 a_\alpha \geq 0\,, \label{an_convex} \\   
  \partial_\theta^2a_\alpha \partial_\theta^4a_\alpha - (\partial_\theta^3a_\alpha)^2 \geq 0\,. \label{RP2} 
\end{align}
The first one implies that the corner function is convex as a function of $\theta$. 
Crucially, the differential inequalities \req{RP} can be ``integrated'' to yield the
lower bounds\cite{Bueno:2016}
\begin{align}
  a_\alpha(\theta) \geq \amin(\theta) \,. 
\end{align} 
As we shall see, the higher the $M$, the stronger the lower bound on $a_\alpha(\theta)$. The lower bound function $\amin(\theta)$
is the solution of the non-linear differential equation obtained by replacing the inequality in \req{RP} by an equality:
\begin{align} \label{amin-ode}
  \det\, \big\{\partial_\theta^{j+k+2} \amin(\theta) \big\}_{j,k=0}^{M} \, = 0\, .   
\end{align}
Thus to obtain $\amin$, we need to solve an initial value problem by providing $M$ initial conditions 
at $\theta=\pi$,\cite{Bueno:2016}  
namely $\partial_\theta^{2s} \amin(\pi)=\partial_\theta^{2s} a_\alpha(\pi)= (2s)!\,\sigma_\alpha^{(s-1)}$,
for $s=1,\ldots,M$. The odd derivatives vanish because of the reflection property, $a_\alpha(\theta)=a_\alpha(2\pi -\theta)$, and $a_\alpha(\pi)=0$. 
We emphasize that this applies to all CFTs. 

The non-linear differential equation (\ref{amin-ode}) for $M\!>\! 0$ generally needs to be solved numerically.
An exception was previously found\cite{Bueno:2016} for $M\!=\! 1$, see \req{RP2}, 
where the exact solution is $\frak a_\alpha^{(1)}(\theta) =\sigma_\alpha\,(\theta-\pi)^2$. 
We note that in this special case the bounding function exactly gives the
corresponding smooth-limit expansion up to order $2M$. This simple feature does not hold at $M>1$. 
Here, we obtain new closed-form expressions for $M=2$:
\begin{align}
  \mathfrak a^{(2)}(\theta) = \frac{\sigma^2}{6\sigma'}\left\{ \cosh\left[ (\tfrac{12\sigma'}{\sigma})^{1/2} (\theta\!-\!\pi) \right] -1 \right\}\,,
\end{align}
and for $M=3$:
\begin{multline}
  \mathfrak a^{(3)}(\theta) = \left(\sigma -\tfrac{2(\sigma')^2}{5\sigma''} \right)(\theta-\pi)^2 \\
  + \frac{2(\sigma')^3}{75(\sigma'')^2} \left\{ \cosh\left[ (\tfrac{30\sigma''}{\sigma'})^{1/2} (\theta\!-\!\pi) \right] -1 \right\}\,,
\end{multline}
where we have omitted the \ren index for clarity. 
It can be explicitly checked that they satisfy \req{amin-ode}. 
As advertised, $\mathfrak a^{(M)}$ is determined by the first $M$
coefficients of the smooth limit expansion of $a(\theta)$. 
Using recently derived bounds\cite{Bueno:2016} for the coefficients $\sigma^{(p)}$, we can explicitly show
that $\mathfrak a^{(3)}(\theta)\geq \mathfrak a^{(2)}(\theta)\geq \mathfrak a^{(1)}(\theta)$. This can be seen
by Taylor expanding the functions about $\pi$. 
Furthermore,
for these 3 values of $M$, the corresponding truncated series at order $(\theta-\pi)^{2M}$ is less than $\mathfrak a^{(M)}$,
and is thus itself a lower bound for $a(\theta)$.
It would be interesting to extend our exact analysis to all values of $M$. However, due to the difficulty in doing so, 
we now proceed via explicit examples. 

\begin{figure}[t]
  \center
  \includegraphics[scale=.36]{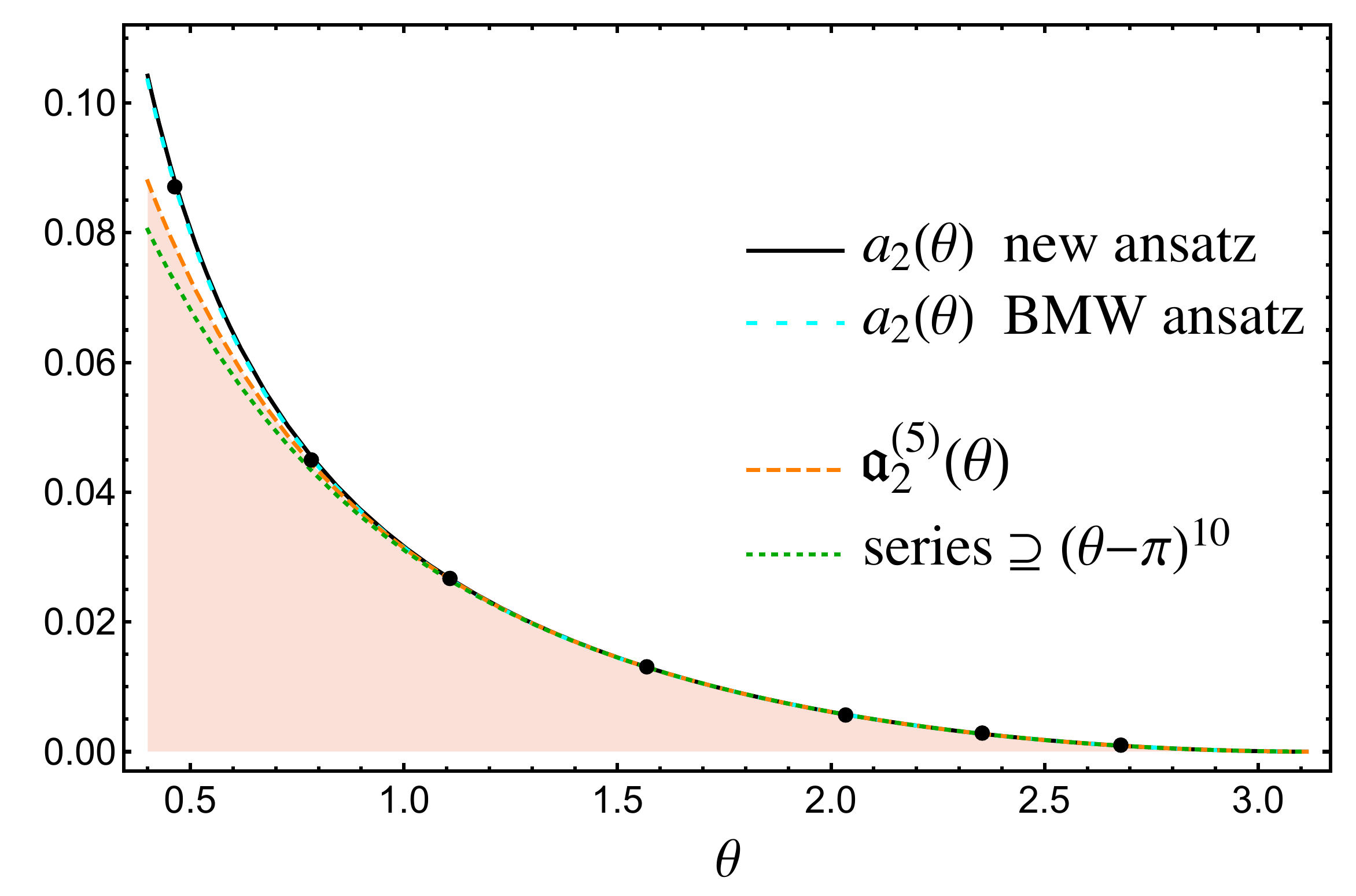} 
  \caption{ Lower bounds on $a_2(\theta)$ for the boson.
  $M\!=\!5$ lower bound $\frak a_2^{(5)}(\theta)$, together
with the corresponding series including terms up to $(\theta-\pi)^{10}$. The shaded region is not allowed.
Also shown are the high precision ansatz \req{new-ansatz},
and the BMW ansatz \req{BMW} for the entire corner function. The markers represent the lattice results, see Section~\ref{sec:latt}. 
  \label{fig:M5}}
  \centering
\end{figure}

We test this new family of lower bounds using the free relativistic boson. 
In Fig.~\ref{fig:M5}, we show the numerical solution for the bounding function at $M\!=\!5$ and
\ren index $\alpha\!=\!2$. Here, $\mathfrak a^{(5)}$ is strictly greater than the $M=1,2,3$ ones given above, hence we do not show the latter.
As a comparison, we also examine the corresponding truncated series to order $(\theta-\pi)^{10}$ (see subsections \ref{sec:qft-b} and \ref{sec:qft-f}),\cite{Casini:2009}
and explicitly confirm that 
$\sum_{p=1}^{5}\sigma_2^{(p-1)}\,(\theta-\pi)^{2p}$ is strictly less than $\frak a_2^{(5)}(\theta)$, see Fig.~\ref{fig:M5}. 
Since the number of terms in the differential equation for $\amin$ grows factorially with $M$, it is quite difficult 
to proceed beyond this $M=5$ bound, at least by naively solving the differential equation. 
Although not shown, analogous results hold for the Dirac fermion. 
Given this evidence, we expect that the following hierarchy will extend to general $M$:
\begin{align} \label{hierarchy}
  \sum_{p=1}^{M}\sigma_\alpha^{(p-1)}\,(\theta-\pi)^{2p} \, \leq\, \amin(\theta) \, \leq\,  a_\alpha(\theta)\,.
\end{align}   
For $2\!<\! M\!\leq\! 5$, we find that the inequalities above are strictly lesser.
In particular, \req{hierarchy} would imply that for any given $M$ we can use the truncated series at order 
$2M$ as a lower bound for the full corner function. 
This latter statement is naturally consistent with the conjecture\cite{Bueno:2016}  
that all the expansion coefficients $\sigma_\alpha^{(p)}$ are non-negative, and that the series about $\theta=\pi$
has radius of convergence $\pi$. All theories for which some or all of the $\sigma_\alpha^{(p)}$ are known satisfy these positivity 
and analyticity properties.\cite{Bueno:2016}  

\subsubsection{Bound for von Neumann}
The inequalities (\ref{RP}) were obtained for the \ren entropies with integer values $\alpha\geq 2$.
Ref.~\onlinecite{Bueno:2016} suggested that they can be extended to all \ren indices $\alpha>0$. We now test and explore the consequences
of this statement for the von Neumann entanglement entropy. In particular, this means that we can set   
$\alpha=1$ in \req{hierarchy}. This provides a new set of lower bounds (one for each $M$), assuming we know
a few of the coefficients $\sigma_1^{(p)}$ for $p\geq 1$. 
These coefficients are known up to $p=7$ for the boson and $p=6$ for the fermion  
(see Ref.~\onlinecite{Casini:2008} for a partial list; 
Tables~\ref{tab:scalar-CFT} and ~\ref{tab:fermion-CFT} in Appendix~\ref{app:smooth_coeffs} give the full list to
high precision). 
As discussed above, $\frak a_1^{(M)}(\theta)$ would in principle provide a better lower bound, 
however it is difficult to obtain numerically at such high values of $M$.  
We thus use the truncated series
to get lower bounds, $a_1(\theta)\geq \sum_{p=1}^M \sigma_1^{(p-1)}(\theta-\pi)^{2p}$,  
for $M=8,7$ respectively. 

These constitute the strongest bounds to date on the corner entanglement at $\alpha=1$ for both theories,
at least away from very small angles. 
As $\theta\to 0$, the true corner function diverges whereas the truncated series
tends to a finite value. For sufficiently small $\theta$, a better bound can be obtained from the strong sub-additivity 
of the entanglement entropy, see \req{ssa-bound}.\cite{Bueno:2016}    
As a further non-trivial check of this new set of bounds, in Appendix~\ref{app:ads} we verify that 
they hold in a class of strongly interacting quantum critical theories described by the gauge/gravity (AdS/CFT) duality.  

We now turn to the explicit field theory calculations for the boson and Dirac fermion theories, for which we obtain new exact results.

\subsection{Massless complex boson}    \label{sec:qft-b}
The Lagrangian density of a massless complex boson, arguably the simplest CFT, is $\tfrac{1}{2}|\partial_\mu \varphi|^2$, 
where $\varphi$ is a complex field. (We have set the velocity to unity.)
Using the replica trick, the corner function 
at integer \ren index $\alpha\!>\! 1$ can be expressed as
(see Ref.~\onlinecite{Casini:2009} and references therein)
\begin{multline} \label{corner-Ren-cs}   
  a_\alpha^{b}(\theta) = \sum_{k=1}^{\alpha-1} \frac{8k(\alpha -k)}{\alpha^2(\alpha-1)}\int_{\tfrac{1}{2}}^\infty\!\! dm \,  
m\sqrt{m^2-\tfrac{1}{4}} \\
\times \int_\theta^\pi\! dy\, H_{\!\frac{k}{\alpha}}(y,m),
\end{multline}
where $H_{\frac{k}{\alpha}}$ determines the trace of the Green's function $\mathcal G_b$ of a massive boson (mass $m$) on a 2-sphere with a cut 
of opening angle $\theta$.\cite{Casini:2007} The cut is simply the intersection of the entangling region (a wedge) with the sphere.
More precisely,
\begin{align}
  \tr \mathcal G_{b} = -8\pi \left(1-\tfrac{k}{\alpha}\right)\tfrac{k}{\alpha}\int_\theta^\pi dy \,  
H_{\!\frac{k}{\alpha}}(y,m)\,. 
\end{align}
For each integer $k$, $k/\alpha$ determines the twist of the Green function across the cut. The form of \req{corner-Ren-cs} follows from 
symmetry:  
even in the presence of the wedge-shaped entangling region,  
the problem has a radial spacetime symmetry. This is why it reduces to  
solving the Green's function on a 2-sphere. 
In the case of the von Neumann entropy $\alpha=1$, the sum in \req{corner-Ren-cs} is replaced by the integral 
\begin{multline} \label{corner-EE-cs}
  a_1^{b}(\theta) = \int_0^\infty\! dt\, \frac{16\pi(t^2+1/4)}{\cosh^2(\pi t)} \int_{\tfrac{1}{2}}^\infty\! dm\, m \sqrt{m^2-\tfrac{1}{4}} \\
  \times \int_\theta^\pi dy\, H_{-it+\tfrac{1}{2}}(y,m)\,. 
\end{multline}

The function $H_a$ is the solution of a complicated set of ordinary non-linear differential equations\cite{Casini:2009}. 
As mentioned above, a convenient way to attack the problem\cite{Casini:2009} is to obtain an expansion for $a_\alpha(\theta)$  
as a series around $\theta=\pi$, \req{series}. We list the first few expansion coefficients in
Table \ref{tab:scalar-CFT} of Appendix~\ref{app:smooth_coeffs}. The leading coefficients $\sigma_\alpha$ are known exactly from previous works\cite{Elvang:2015,Bueno:2015}.
We have obtained new exact answers for $\sigma_\alpha'$, with $\alpha=1,2,3$. For instance, $\sigma_1'=(20+3\pi^2)/(9216\pi^2)\approx 
5.45\times 10^{-4}$; 
the derivation of this new result is outlined in Appendix~\ref{app:sigma1}.   
Ref.~\onlinecite{Bueno:2016} conjectured that $\sigma_1'$   
is related to the groundstate 4-point function of the stress tensor; it would be interesting to test this claim for the free boson using
our exact result. In this respect, the methods of Refs.~\onlinecite{Faulkner:2015aa,Bianchi:2015}
could be used.  
The other coefficients $\sigma_\alpha^{(p>1)}$ were obtained by numerically evaluating integrals to high precision.   
\begin{figure}  
  \center 
  \includegraphics[scale=.415]{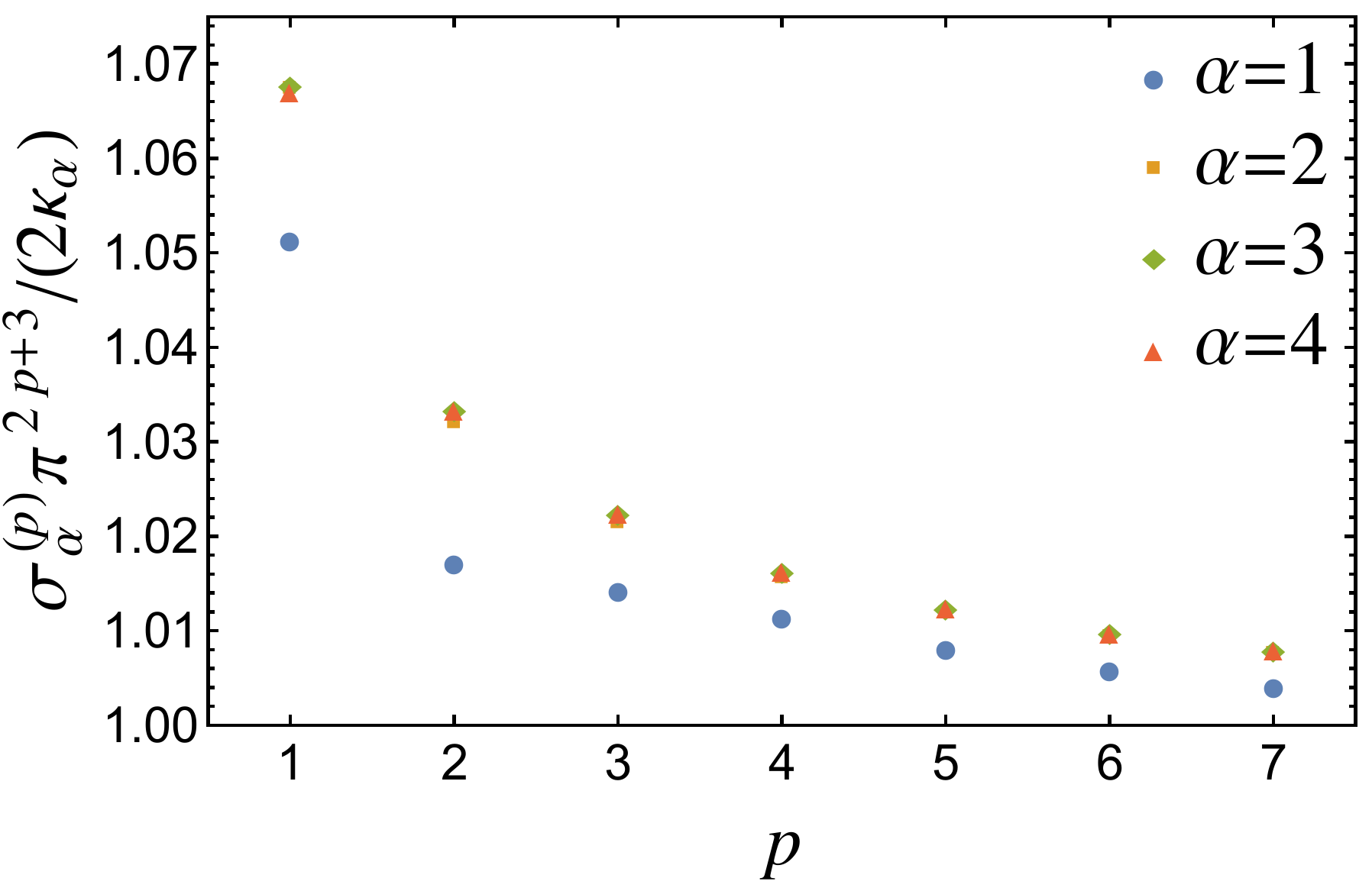}  
  \caption{$\sigma_\alpha^{(p)}$ coefficients for the boson, see Table~\ref{tab:scalar-CFT}, normalized by their large-$p$ asymptotic value, $2\kappa_\alpha / \pi^{2p+3}$.
The $p\!=\!0$ ones are not shown because they are substantially greater. The $\kappa_\alpha$ are given in Ref.~\onlinecite{Bueno:2015JHEP}.}    
  \label{fig:sig_b}  
  \centering    
\end{figure}

\begin{figure} 
  \center 
  \includegraphics[scale=.35]{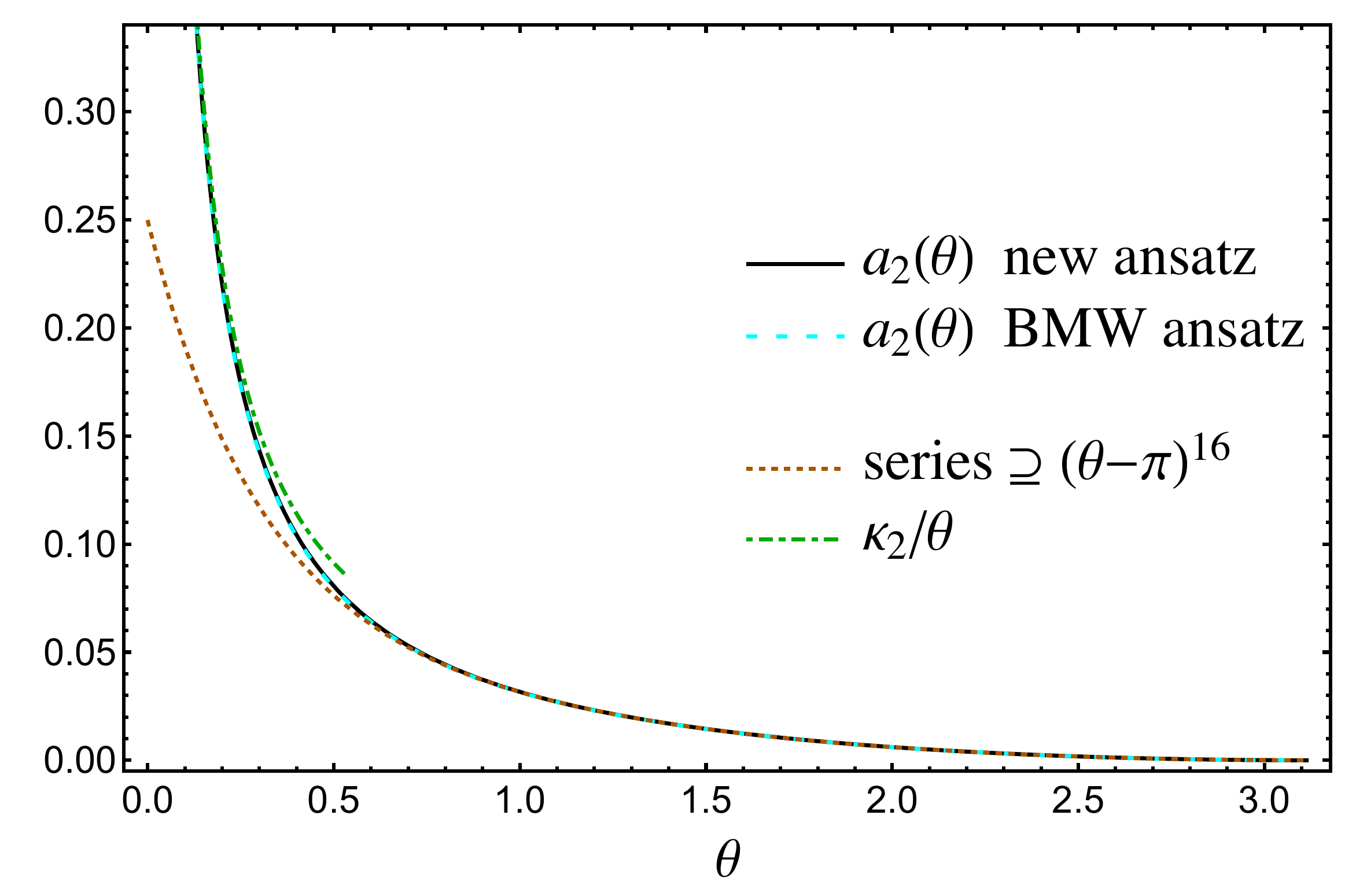}
  \caption{Comparison of the field theory ansatze with the
truncated series for $a_2^{b}(\theta)$ including terms up to order $(\theta-\pi)^{16}$. The series is expected to be a lower bound. 
Also shown is the small angle behavior $\kappa_2/\theta$.} 
  \label{fig:sigmap}  
  \centering    
\end{figure}

\subsection{Massless Dirac fermion}  \label{sec:qft-f}
A single Dirac fermion has Lagrangian density $\bar\psi i\gamma_\mu\partial_\mu\psi$, where $\psi$ is
a 2-component spinor.
The calculations of the entanglement entropies are analogous to the boson case described above, and we refer the reader
to Ref.~\onlinecite{Casini:2009} for the details. 
We list the first 7 smooth-limit expansion coefficients in
Table \ref{tab:fermion-CFT} of Appendix~\ref{app:smooth_coeffs}. The leading coefficients $\sigma_\alpha$ are known exactly from previous works.\cite{Bueno:2015,Elvang:2015,Bueno:2015JHEP}  
As for the boson, we have obtained new exact answers for $\sigma_\alpha'$, $\alpha=1,2,3$. For example, 
we have found $\sigma_1'=(16+3\pi^2)/(9216\pi^2)\approx 5.01\times 10^{-4}$, which is less than the bosonic answer, 
but nevertheless shares a very similar structure with it. 
The other $\sigma_\alpha^{(p>1)}$
were obtained by numerically evaluating integrals with high precision.   
 
\subsection{High precision ansatz}
We now discuss ways to obtain high-precision estimates for the corner entanglement function at all angles. This
becomes especially valuable at small angles, such as $\theta\! =\! \arctan(1/2)$, because there the truncated series 
(to the order we have computed it) cease to be accurate.  
Ref.~\onlinecite{Bueno:2015JHEP} proposed a simple closed-form ansatz for the corner function $a_\alpha(\theta)$ given 
in terms of the smooth limit coefficient $\sigma_\alpha$  
and the sharp limit one, $\kappa_\alpha$: 
\begin{multline}  \label{BMW}
  a_\alpha^{\rm BMW}(\theta) =  \frac{2\pi(\kappa_\alpha-3\pi \sigma_\alpha)}{\pi^2-6}\frac{(\theta-\pi)^2}{\theta(2\pi-\theta)} \\ 
  - \frac{3(2\kappa_\alpha-\pi^3 \sigma_\alpha)}{\pi(\pi^2 - 6)}\left[ 1+(\pi-\theta)\cot\theta \right].  
\end{multline}
We refer to this equation as the ``BMW ansatz'' below.
This function obeys the exact asymptotics as $\theta\!\to\! 0,\pi$, 
and satisfies non-trivial properties required for the corner function.\cite{Bueno:2015JHEP}  
In fact, it is a linear combination of corner functions for 1) a class of Lifshitz quantum critical points,\cite{Fradkin:2006} 
and 2) the so-called Extensive Mutual Information model,\cite{Casini:2009JHEP,Swingle:2010} respectively.     

Here we derive a more refined ansatz for $a_\alpha(\theta)$ by employing the full truncated series instead of only the first 
term, proportional to $\sigma_\alpha$. This ``new ansatz'' takes the form: 
\begin{multline}  \label{new-ansatz}
  \tilde a_\alpha(\theta) =\! \sum_{p=1}^M \sigma_\alpha^{(p-1)}\, (\theta-\pi)^{2p}+ \frac{2\kappa_\alpha}{\pi^{2M+1}} 
  \frac{(\theta-\pi)^{2(M+1)}}{\theta (2\pi-\theta)}
\end{multline} 
where $M$ corresponds to the number smooth limit coefficients used. Whenever we quote numerical values using
this ansatz we use $M=7,8$ for the fermion and boson, respectively. 
In \req{new-ansatz}, the finite sum is simply the smooth
limit expansion up to order $2M$. 
The second term follows from exactly performing the infinite
sum, $\sum_{p=M+1}^\infty \sigma_\alpha^{(p-1)}(\theta-\pi)^{2p}$, with the 
replacement $\sigma_\alpha^{(p-1)}\to 2\kappa_\alpha/ \pi^{2p+1}$. The series is geometric and can be simply evaluated.
The motivation behind this replacement is the observation that for sufficiently large $p$, the coefficients decay exponentially as\cite{Bueno:2016}
\begin{align} \label{sig-asym}
  \sigma_\alpha^{(p)} \simeq \frac{2\kappa_\alpha}{\pi^{2p+3}}\,, \quad p\gg 1. 
\end{align}
This ``smooth-sharp connection'' follows from the assumption that the series about $\theta=\pi$ has radius of convergence $\pi$: thus, summing all the high $p$
terms must yield a $\kappa_\alpha/\theta$ divergence at small angles, which yields the asymptotics \req{sig-asym}. 
The new ansatz thus also takes as input the value of the sharp limit coefficient $\kappa_\alpha$.   
\req{new-ansatz} will thus have the correct $M$ first coefficients $\sigma_\alpha^{(p)}$, and 
the correct behavior as $\theta\to 0$. In contrast the BMW ansatz \req{BMW} 
only takes as input the first coefficient $\sigma_\alpha$ (and $\kappa_\alpha$): it will not be as accurate as the new ansatz 
for large angles.
We note that the BMW ansatz also captures the correct small angle dependence. 
As illustrated using $a_2(\theta)$ for the boson in Fig.~\ref{fig:M5}, 
both the new ansatz and the BMW one yield very similar results. This being said, \req{new-ansatz} will be the more accurate of the two, and to our knowledge constitutes the
most precise estimate for $a_\alpha(\theta)$. Indeed, the validity of the ansatz for the free boson and fermion CFTs is supported 
by the fact that for $M=7,8$,
the smooth limit coefficients are already very close to their asymptotic form, \req{sig-asym}, as illustrated in 
Fig.~\ref{fig:sig_b}. In that figure we further see that the $\sigma^{(p)}$ approach their asymptotic
value from above, suggesting that our new ansatz is also a lower bound for $a_\alpha(\theta)$. 

\section{\Renyi entropies in free field theories on the lattice} \label{sec:latt} 
In this section, we review the basics of the exact computation of \Renyi entropies on finite lattices in the free boson and the free Dirac fermion theories.
We use correlator-based methods \cite{Bombelli1986,Srednicki1993,Peschel} that rely on Wick's theorem being satisfied in the ground state.
Specifically, we numerically evaluate two-point correlators restricted to the sub-region $A$ in the ground state and diagonalize the resulting matrices to obtain $S_\alpha(A)$. 
Numerical methods for evaluating these correlators are a useful alternative to analytical integration techniques in the continuum\cite{Casini:2007,Casini:2008}, particularly in higher space-time dimensions, where high-precision analytical calculations may become infeasible (or impossible).
We perform our calculations on the square lattice with a finite number of lattice sites $N$, motivated by the setup for interacting theories in two dimensions \cite{Kallin:2013,Kallin:2014,Miles:2014}.
Similar calculations have previously been performed at special angles in Ref.~\onlinecite{Casini:2007,Casini:2009}.

\subsection{Free Boson} \label{sec:free_boson_EE} 
Consider a finite two-dimensional square lattice, such that a complex scalar field $\phi_{x,y}$ and its conjugate momentum $\pi_{x,y}$ exist at each lattice point.
The free theory is described by the following lattice Hamiltonian,
\begin{align}\label{eq:bosonhamiltonian}
H = \frac{1}{2} \sum _{x,y =1,1}^{L_{x},L_{y}} \Big[ & |\pi_{x,y}|^{2} +  |\phi_{x+1,y} - \phi_{x,y}|^{2}  \\
&+ |\phi_{x,y+1} - \phi_{x,y}|^{2} + \;m^{2} |\phi _{x,y}| ^{2}\Big],\nonumber
\end{align}
where $L_x$ and $L_y$ are the linear dimensions of the lattice with open boundary conditions and $|\mathcal{O}|^2 = \mathcal{O}^\dagger \mathcal{O}$ for operator $\mathcal{O}$. Simplifying the notation with the total number of sites $N=L_x L_y$ and $i = x L_y + y$, the Hamiltonian may be rewritten as
\begin{equation}
 H=\frac{1}{2} \sum_{i=1}^{N} \pi_i^\dagger \pi_i + \frac{1}{2}\sum_{i,j=1}^N \phi^\dagger_i M_{ij} \phi_j.
\end{equation}
The matrix $M$ is the discrete laplacian (up to a diagonal term) on the square lattice.  
The groundstate two-point correlations are given by $X_{ij}=\langle \phi^\dagger_i \phi_j \rangle = (M^{-1/2})_{ij}$ and $ P_{ij}=\langle\pi^\dagger_i \pi_j\rangle = (M^{1/2})_{ij}$.
These correlations and the resulting matrices $X$ and $P$ may be obtained explicitly in many geometries and numerically in others.
See, for example, Ref~\onlinecite{Sahoo} for explicit expressions for $\langle \phi^\dagger_i \phi_j \rangle$ and $\langle\pi^\dagger_i \pi_j\rangle$ for both Dirichlet open boundary conditions (with $\phi_i=0$ outside the lattice) and periodic boundary conditions.

In order to calculate the entanglement entropies, we use the method first introduced in Ref.~\onlinecite{Peschel}.
The two-point correlators define the reduced correlation matrix $C_A = \sqrt{X_A P_A}$, where $X_A$ and $P_A$ are the matrices $X$ and $P$ (respectively), but with indices $i$, $j$ restricted to region $A$.
The von Neumann and \Renyi entropies for the free boson Hamiltonian of Eq.~\eqref{eq:bosonhamiltonian} can be written in terms of the eigenvalues $\lambda_i$ of the matrix $C_A$ as\cite{Casini:2009}
\begin{align}
S_1(A) &= \sum_{i=1}^{N_A} \left[ \left( \lambda_i + 1 \right) \ln \left( \lambda_i + 1 \right) \right. \nonumber \\
& \phantom{= \sum [}- \left. \left( \lambda_i - 1 \right) \ln \left( \lambda_i - 1 \right)  - 2\log(2) \right], \label{eq:boson_EE_vN}\\
S_\alpha(A) &= \frac{2}{\alpha-1} \sum_{i=1}^{N_A} \ln \left[   \frac{(\lambda_i+1)^\alpha - (\lambda_i-1)^\alpha}{2^\alpha} \right], 
\label{eq:bosonentropy}
\end{align}
where $N_A$ is the number of sites $A$. 

We note that in practice the numerical calculations are performed for the \emph{real} scalar boson as opposed to the complex one.
We then multiply all the entropies by two to extract the corner coefficient of the complex boson to compare to the field theory calculations of 
previous sections.

\subsection{Free Fermion}	\label{sec:free_fermion_EE}

Consider a square lattice with two fermions per site. In order to obtain a single Dirac fermion at low energy, we assume that the Hamiltonian in momentum space has the form
\begin{equation}
H = \sum_{\vec{k}}  \psi^{\dagger}_{\vec{k}}\, ( \vec{h}(\vec{k})\cdot \bm{\sigma} ) \, \psi_{\vec{k}} , \label{eq:ChernInsulatorHam}
\end{equation}
with
\begin{equation}
\vec{h}(\vec{k}) = \left(\sin k_x , \sin k_y, 2-\cos k_x - \cos k_y \right) , 
\label{eq:hk}
\end{equation}
where $\Psi_{\vec{k}}$ is the fermion spinor at momentum $\vec{k}=(k_x, k_y)$, $\bm{\sigma}$ is the vector of Pauli matrices and we have chosen units in which the overall energy scale is one.
We assume periodic boundary conditions along the $x$-direction and anti-periodic boundary conditions along the $y$-direction; the anti-periodic boundary conditions avoid the zero mode at $k_x=k_y=0$ at finite system size $N$.
Observe that as $k_x, k_y \rightarrow 0$, $H(\vec{k}) \approx k_x \sigma_x + k_y \sigma_y$, which is the Hamiltonian corresponding to a single Dirac fermion. As $H$ breaks time-reversal symmetry, the fermion doubling theorem does not apply.

We again use the method discussed in Ref.~\onlinecite{Peschel} which relates the eigenvalues of the reduced density matrix $\rho_A$
to those of a ground state correlation matrix.
In this case, the latter is given by $C_{ij} = \langle c_i^\dagger c_i \rangle$ for lattice sites $i$ and $j$.
We restrict $C$ to the region $A$ and denote the corresponding matrix by $C_A$.
The expressions for the von Neumann and \Renyi entropies in terms of the eigenvalues $\lambda_i$ of $C_A$ read
\begin{eqnarray}
S_1(A)& =& \sum_{i=1}^{2N_A} -\lambda_i \ln \lambda_i - (1- \lambda_i) \ln (1- \lambda_i) , \\
S_\alpha(A)& =& \frac1{1-\alpha} \sum_{i=1}^{2N_A} \ln \left( \lambda_i^\alpha + (1-\lambda_i)^\alpha \right).
\label{eq:fermionentropy}
\end{eqnarray}

\subsection{Numerical extraction of corner terms} \label{NumExt}

In order to obtain $a_{\alpha}(\theta)$ from Eqs.~\eqref{eq:bosonentropy} and \eqref{eq:fermionentropy}, the leading area law term in Eq.~\eqref{eq:S_corner} must be removed either with fitting or by subtraction.
The length scales involved in the calculation are the linear size of the entire system $L$  
(infrared regulator of the theory), the lattice length scale $\delta$ (ultraviolet regulator), and the bipartition length $\ell$.  
Eq.~\eqref{eq:S_corner} holds in the regime $L \gg \ell \gg \delta$. 

We employ two different strategies to obtain $a_\alpha(\theta)$ in the thermodynamic limit.
The first is the \emph{direct approach}, in which we compute the entropy of a region $A$ 
defined by a regular polygon with $n$ sides of length $\ell$, embedded in an $N = L \times L$ torus.
The simplest scenario for this type of calculation is for region $A$ to be a square, 
with four corners.  
In order to suppress any ratio-dependent sub-leading contributions to Eq.~\eqref{eq:S_corner}, 
we fix the aspect ratio of the square and torus, e.g.~$ L / \ell =4$ as used below.
We then produce data for a range of sufficiently large $\ell$ in order to extract $n a_\alpha(\theta)$ as the coefficient of the logarithm in Eq.~\eqref{eq:S_corner}.
For the square with $n=4$ corners for example, we obtain $4a_\alpha(\pi/2)$ as the coefficient.

\begin{figure}[t]
    \centering
    \includegraphics[width=0.45\linewidth]{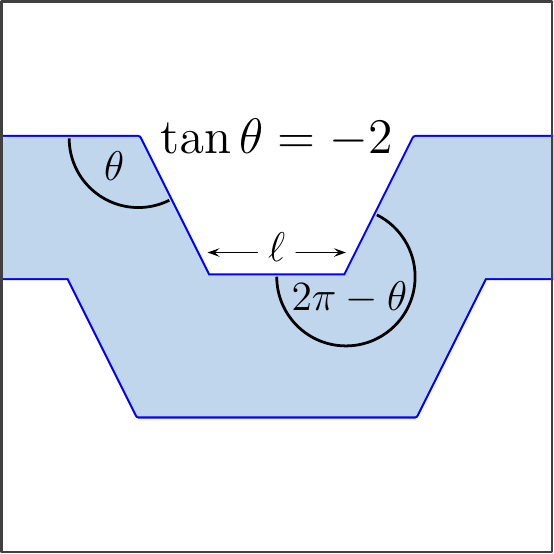}
    \hspace{0.5cm}\includegraphics[width=0.45\linewidth]{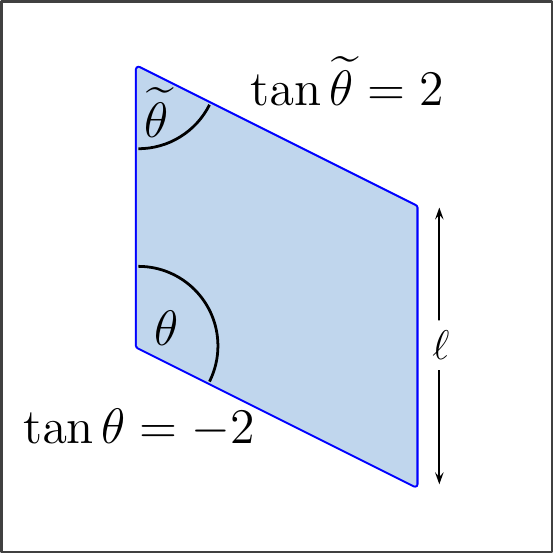}
\caption{Closed polygons in periodic boundary conditions in order to obtain the corner contribution from fitting the area law to the full entanglement entropy. Examples are shown for $\tan \theta = -2$ (left panel) and for $\tan \widetilde{\theta} = 2$ (right panel) which requires the subtraction the $\theta$-contribution.}
    \label{fig:polygoncorners}
\end{figure}

The same approach provides the corner coefficient for angles other than $\theta = \pi /2$ if we choose region $A$ appropriately.
For example, in the left panel of Fig.~\ref{fig:polygoncorners}, $A$ is a closed polygon that winds around the torus with two trapezoidal pits whose corners subtend angles $\theta$ and $2\pi - \theta$. 
As the corner coefficient is symmetric (Eq.~\eqref{Eq:asymmetric}), the net coefficient of the logarithm for $A$ in the left panel of Fig.~\ref{fig:polygoncorners} is $ 8 a_\alpha (\theta) $.
Computing the corner coefficient for angles $ \widetilde{\theta} < \pi/2$ involves the subtraction of at least one other previously-obtained corner contribution. 
E.g.~for $\pi / 4$ we take an isosceles right-angled triangle and subtract the $\pi /2 $ corner contribution.
Similarly we use a parallelogram with angles such that $ \tan \widetilde{\theta} = 2 $ and $\tan \theta = -2 $ (Fig. \ref{fig:polygoncorners} right panel) and obtain the coefficient $a_\alpha(\widetilde{\theta})$ by subtracting the contribution from the angle $ \theta = \pi - \widetilde{\theta}$ to the logarithmic coefficient. 

In the direct approach, we include sub-leading contributions to the functional form of Eq.~\eqref{eq:S_corner} in order to obtain a reliable fit. As discussed in the next section, this is particularly problematic for the free boson, due to the presence of a $\mathcal{O}(1/L)$ sub-leading correction.
In contrast, it is successful for the free fermion, which has much weaker sub-leading corrections, see subsection~\ref{IR}.  
The periodic boundary conditions of the torus enable us to access system sizes of $\sim 10^4$ sites for subsystem $A$. 

\begin{figure*}[t]
    \centering
    \includegraphics[width=0.23\linewidth]{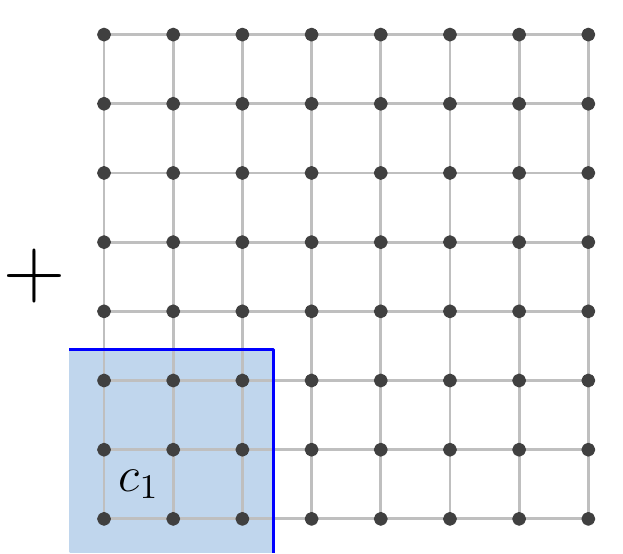}
    \hspace{0.0cm}\includegraphics[width=0.23\linewidth]{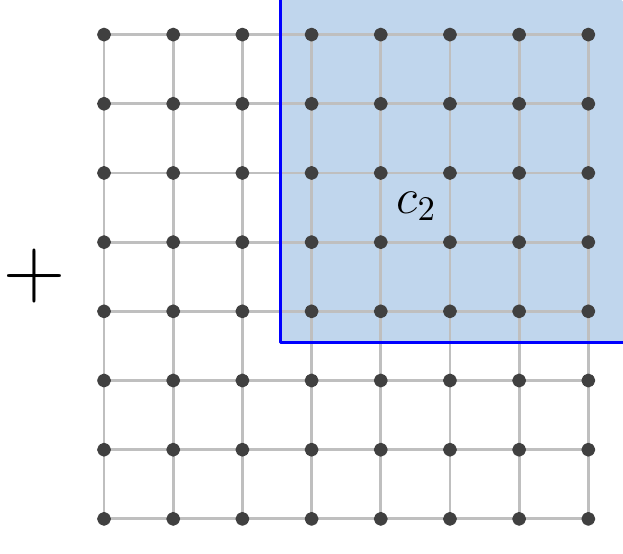}
    \hspace{0.0cm}\includegraphics[width=0.23\linewidth]{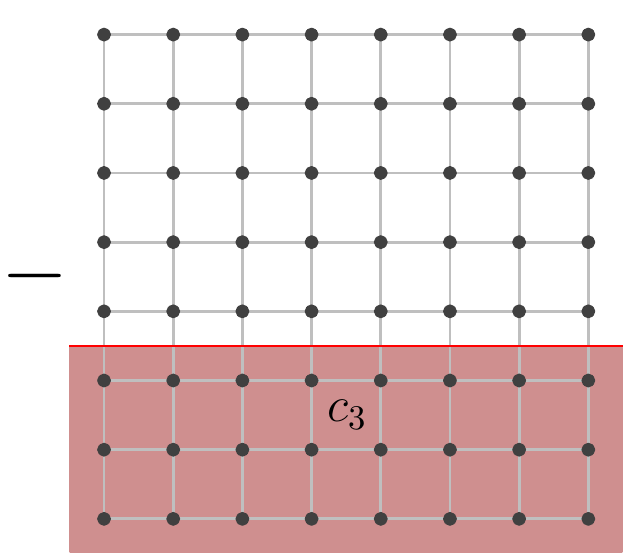}
    \hspace{0.0cm}\includegraphics[width=0.23\linewidth]{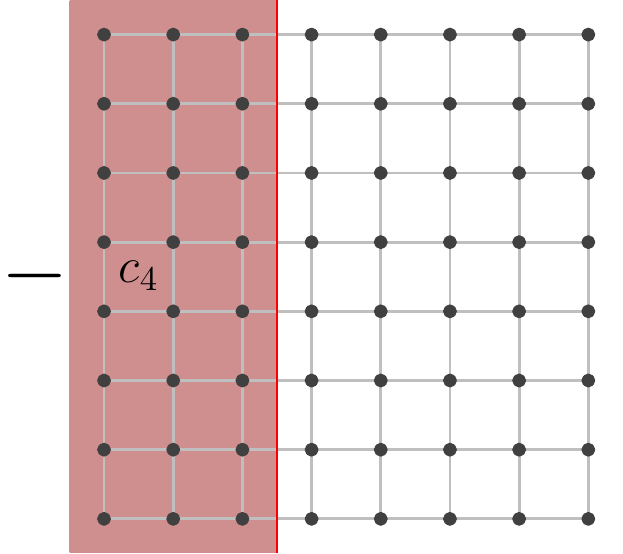}
    \caption{Bipartitions $c_1$, $c_2$, $c_3$ and $c_4$ used to determine the corner entanglement for $\theta = \pi / 2$ according to Eq.~\eqref{eq:propertycalc}.}
    \label{fig:cornerbipartitions}
\end{figure*}

An alternative strategy to the direct approach \cite{Kallin:2013,Kallin:2014,Miles:2014,Sahoo} is to perform finite-size scaling with the numerical linked-cluster expansion (NLCE).
In addition to systematically accessing the limit $L,\ell \rightarrow \infty$, the NLCE eliminates the area-law piece and $\mathcal{O}(1/L)$ corrections to the \Renyi entropy, providing direct access to the corner coefficient.  
The NLCE is based on determining a certain ``property'' $P$ of interest for all embeddable subclusters $c$ of the lattice up to the maximally achievable linear size, dubbed cluster order. 
We use the variant developed in Ref.~\onlinecite{Kallin:2014} and restrict ourselves to open-boundary rectangular clusters.
Every cluster is assigned a weight 
\begin{align}
    W(c) = P(c) - \sum_{s \in c} W(s),
\end{align}
where $s$ is a subcluster of $c$, such that the ``property'' (defined below) divided by the number of cluster sites $N$ for the total lattice of linear size $L$ is
\begin{align}
    \frac{P(L)}{N} = \sum_c e(c) \times W(c),
\end{align}
with an embedding factor $e(c)$ that is one for square clusters and two for rectangular clusters with differing length of both dimensions.\cite{Kallin:2013}

Our ``property'' $P$ is the sub-leading logarithmic term of Eq.~\eqref{eq:S_corner} 
which can be elegantly obtained by computing the entropies for four bipartitions $c_1$, $c_2$, $c_3$ and $c_4$ (depicted for a corner angle $\theta = \pi / 2$ in Fig.~\ref{fig:cornerbipartitions}).
Bipartitions $c_1$ and $c_2$ both possess a single corner while $c_3$ and $c_4$ are corner-free but have the same total boundary length. Hence, adding them as \begin{align}
    P_r(c) = \frac12 \left [ S(c_1) + S(c_2) - S(c_3) - S(c_4)  \right ],
    \label{eq:propertycalc}
\end{align}
intrinsically cancels the leading boundary contribution.
This procedure is repeated for every possible location $r$ of the corner and the results are added;
$P(c) = \sum_r P_r(c)$.
Finally, we perform a fit of the total property $P(L_c)$ against $\ln L_c$ -- where $L_c$ is a measure of the cluster length scale\cite{Sahoo} -- in order to obtain $a_\alpha(\theta)$.

To access the corner coefficient of angles other than $\theta = \pi /2 $ in NLCE, we pixelate (at the ultraviolet scale) one of the lines whose intersection defines the vertex.
This is shown in Fig.~\ref{fig:corner45} for $\theta=\pi/4$ and $\theta = \arctan(-2)$. 
The pixelation gives access to the family of opening angles $\theta = \arctan(p/q)$ for integer $p,q$. 

The final values for $a_\alpha(\theta)$ are obtained from the following two-step extrapolation.
The initial raw data are the values of the \emph{property} from the NLCE or the direct approach.
First, we perform a linear fit of these values against $\ln L$ (least square fit to $\mathcal{A}_\alpha \ell - a_\alpha \ln \ell + c$)  in order to extract the corner coefficient in the NLCE (direct approach).
We take fitting intervals of $L$ of constant size $\Delta L$, i.e. $[ L_\text{max} - \Delta L, L_\text{max} ]$, and obtain corresponding $a_\alpha^{(L_\text{max})}(\theta)$.
Second, we carry out an extrapolation of these values to the thermodynamic limit $L_\text{max} \rightarrow \infty$.
This is done allowing for an arbitrary scaling exponent $p$, i.e. fit $a_\alpha^{(L_\text{max})}(\theta)$ to $A + B / L_\text{max}^p$ 
as in Ref. \onlinecite{Sahoo} and taking $A$ as the final value for $a_\alpha(\theta)$.
For the boson, we achieve cluster sizes up to $L = 55$.

The direct approach and the NLCE algorithm are complementary methods for extracting the corner coefficient and performing systematic finite size scaling on any lattice model.
A close examination of the convergence properties of each is generally necessary to
determine which method is the most suitable for any given model.
We find that the NLCE is very successful at performing reliable finite size scaling for the free boson, as it eliminates the $\mathcal{O}(1/L)$ correction to the \Renyi entropy discussed in Section \ref{IR}.
In contrast, we find that obtaining sufficient data to perform the required fits with NLCE 
is challenging for the free fermion.  
There, one has two sites in each unit cell, which doubles the linear size of every matrix involved, exacerbating finite size effects.
Thus, in the results below, we generally use NLCE to obtain boson data, and the direct approach for the Dirac fermion.

\subsection{Results} 

In this section, we use the numerical techniques outlined above to compute $ a_\alpha(\theta)$ for the free boson and the free Dirac fermion.
In Section \ref{IR}, we present the corrections to $S_\alpha(A)$ due to the finite infra-red regulator $L$.  
These corrections affect the extrapolation to the thermodynamic limit, as discussed in the previous section. 
Next, in \ref{Sec:AngleDependenceCorner} we analyze our results for $a_\alpha^{b,f}(\theta)$ and compare them with our new field theory estimates,
and with the literature. 
In Section \ref{BF}, we use our best numerical extrapolation techniques to examine the duality between the entropy corner coefficient for free bosons and fermions.

\subsubsection{Infra-red scaling} \label{IR}

We examine the infra-red (IR) contribution to the entropy for the free boson Hamiltonian of Eq.~\eqref{eq:bosonhamiltonian} and the free fermion Hamiltonian of Eq.~\eqref{eq:ChernInsulatorHam}. 
To isolate this IR contribution, we perform calculations of the entanglement entropy $S_1(A)$ in the case where region $A$ is a fixed size, so that all contributions that scale as $\ell/\delta$ are constant. 
Below we take $A$ to be a single site, although the exercise could be repeated for general regions.
As we discussed in Secs.~\ref{sec:free_boson_EE} and~\ref{sec:free_fermion_EE}, the entropies are calculated from the two-point correlation functions inside of region $A$.
When $A$ is a single site, we thus only need to calculate the correlators on this one site.

\begin{figure}[t]
    \centering
    \includegraphics{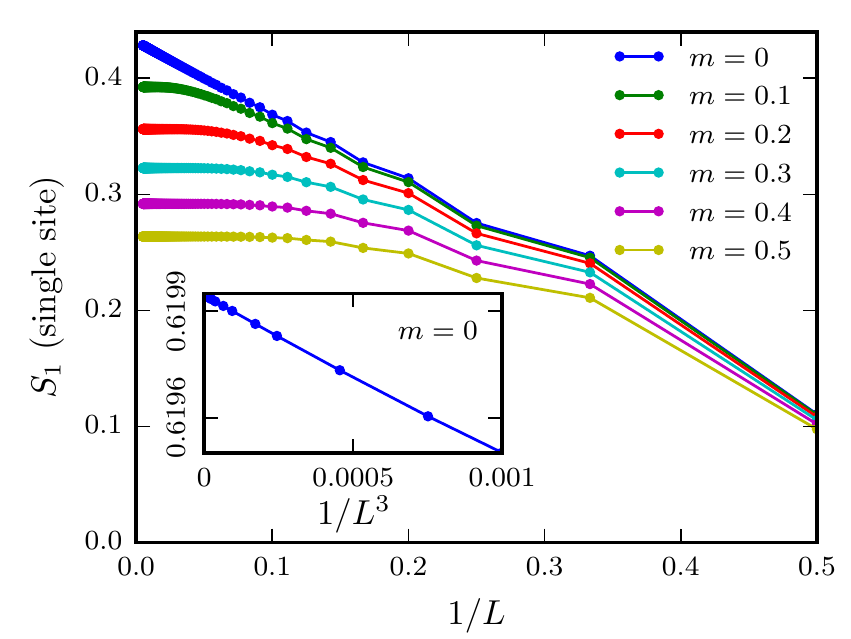}
    \caption{The entropy $S_1$ for free bosons as a function of $1/L$ for the case where region $A$ is a single site and the lattice has open boundary conditions (the same as the boundary conditions used in the NLCE). As expected, we see that the leading IR contribution to the entropy is $\mathcal{O}(1/L)$ when the boson is massless ($m=0$). 
    Similarly, the inset shows the single-site entropy for the massless fermion (on a lattice with antiperiodic boundary conditions) and illustrates that the leading IR contribution is $\mathcal{O}(1/L^3)$, which corresponds to a faster decay than the entropy for the free massless boson.
    }
    \label{fig:IR_bosonsFermions}
\end{figure}

For massless free bosons, the correlators for a single-site region $A$ are given in the (non-discretized) 
field theory by 
$\langle \phi_i^\dagger \phi_i \rangle \!\sim\! \int \text{d}^2\mathbf{k}\, \frac{1}{k}$ 
and 
$\langle \pi_i^\dagger \pi_i \rangle \!\sim\! \int \text{d}^2\mathbf{k} \, k $. 
Since the magnitude $k$ must be integrated from $2\pi/L$ to $2\pi/\delta$, one finds that the IR contribution is $\mathcal{O}(1/L)$ for $\langle \phi_i^\dagger \phi_i \rangle$ and $\mathcal{O}(1/L^3)$ for $\langle \pi_i^\dagger \pi_i \rangle$.
Substituting into Eqs.~\eqref{eq:boson_EE_vN} and~\eqref{eq:bosonentropy} then gives that the leading IR contribution to the entropy $S_\alpha$ is $\mathcal{O}(1/L)$.
In Fig.~\ref{fig:IR_bosonsFermions}, we show numerically that the entropy $S_1$ does indeed become linear in $1/L$ as the boson becomes massless ($m \to 0$).
This figure shows results for open boundary conditions, and we also observe the same scaling behavior on lattices with fully anti-periodic boundary conditions or with periodic boundary conditions in one direction. Note that the entropy is divergent when the lattice has fully periodic boundary conditions and $m=0$ due to the zero mode at $k_x = k_y = 0$.
In general spatial dimension $d \geq 2$, one can show similarly (and verify numerically) that the leading IR contribution is $\mathcal{O}(1/L^{d-1})$.
One generally expects, and we have confirmed numerically, that this IR scaling survives for regions $A$ of fixed arbitrary shape in the limit $L \rightarrow \infty$.  
The finite size of $A$ merely affects the crossover value of $L$ beyond which the $1/L$ scaling holds in Fig.~\ref{fig:IR_bosonsFermions}.

For massless fermions, similar calculations reveal that the two-point correlators scale as $\mathcal{O}(1/L^3)$ and the leading IR contribution to the entropy is also $\mathcal{O}(1/L^3)$.
This behavior is observed in our numerical calculations and illustrated in the inset of Fig.~\ref{fig:IR_bosonsFermions}.
The rapid decay of the leading IR correction means that finite size effects
are much less pronounced for fermions than bosons.

\subsubsection{Angle dependence of corner entanglement} 
\label{Sec:AngleDependenceCorner}

We use the methods of Sec.~\ref{NumExt} to determine the corner contribution for general angles $\theta$.
In addition to $\theta = \pi/2$, we choose six other angles, namely those which can be pixelated most naturally on a square lattice, 
i.e.~where the slope of one line is $ \pm \frac12, \pm 1, \pm 2 $, as illustrated in Fig.~\ref{fig:corner45}.
The actual angle is obtained by taking the arctangent of these slopes and transposing it into the range of $[0, \pi]$.
The finite angle results are given in Tables \ref{tab:boson} and \ref{tab:fermion} for the free boson and Dirac fermion, respectively.
These tables also include field theory estimates, discussed in detail in Section~\ref{sec:QFT}.  

\begin{table*}   
  \centering
  \begin{tabular}{c||c|c|c||c|c|c||c|c|c||c|c|c}
    \multirow{2}{*}{$\tan\theta$} & \multicolumn{3}{c}{$\alpha=1$} & \multicolumn{3}{c}{$\alpha=2$} & \multicolumn{3}{c}{$\alpha=3$} & \multicolumn{3}{c}{$\alpha=4$} \\ 
                               & {\bf series} & {\bf ansatz} & {\bf lattice} & {\bf series} & {\bf ansatz} & {\bf lattice} & {\bf series} & {\bf ansatz} & {\bf lattice} & {\bf series} & {\bf ansatz} & {\bf lattice} \\ \hline 
        1/2 & 0.1453  & 0.156  & 0.154  & 0.08208 & 0.0881 & 0.087  & 0.06686 & 0.0718 & 0.069 & 0.06041 & 0.0649 & 0.061  \\ \hline
          1 & 0.08037 & 0.0810 & 0.0809 & 0.04494 & 0.0453 & 0.0450 & 0.03652 & 0.0368 & 0.0365 & 0.03296 & 0.0332 & 0.033   \\ \hline
          2 & 0.04816 & 0.0482 & 0.0483 & 0.02668 & 0.0267 & 0.0268 & 0.02163 & 0.0217 & 0.0215 & 0.01952 & 0.01953 & 0.019 \\ \hline
   $\infty$ & 0.02367 & 0.0237 & 0.0236 & 0.01297 & 0.0130 & 0.0130 & 0.01049 & 0.01049 & 0.0104 & 0.009459 & 0.009459 & 0.0092 \\ \hline
         -2 & 0.01051 & 0.0105 & 0.0105 & 0.005718 & 0.00572 & 0.00572 & 0.004617 & 0.004617 & 0.00454 & 0.004160 & 0.004160 & 0.0040 \\ \hline  
         -1 & 0.005040 & 0.00504 & 0.00507 & 0.002733 & 0.00273 & 0.00272 & 0.002205 & 0.002205 & 0.00218 & 0.001986 & 0.001986 & 0.0019 \\ \hline  
       -1/2 & 0.001705 & 0.00171 & 0.00170 & 0.0009226 & 0.000923 & 0.000923 & 0.0007439 & 0.0007439 & 0.000747 & 0.0006700 & 0.0006700 & 0.00067
  \end{tabular}      
  \caption{{\bf Boson:} Finite angle results for $a_\alpha(\theta)$ for the boson at \ren index $\alpha=1,2,3,4$. 
The series and ansatz \req{new-ansatz} results
are obtained using field theory, whereas the lattice results are obtained numerically using NLCE. 
The truncated series  
result includes terms up to $(\theta-\pi)^{16}$, and can be taken as a lower bound for $a_\alpha(\theta)$.}  
  \label{tab:boson}     
\end{table*}  

\begin{table*}    
  \centering
  \begin{tabular}{c||c|c|c||c|c|c||c|c|c||c|c|c}
    \multirow{2}{*}{$\tan\theta$} & \multicolumn{3}{c}{$\alpha=1$} & \multicolumn{3}{c}{$\alpha=2$} & \multicolumn{3}{c}{$\alpha=3$} & \multicolumn{3}{c}{$\alpha=4$} \\ 
                               & {\bf series} & {\bf ansatz} & {\bf lattice} & {\bf series} & {\bf ansatz} & {\bf lattice} & {\bf series} & {\bf ansatz} & {\bf lattice} & {\bf series} & {\bf ansatz} & {\bf lattice} \\ \hline 
        1/2 & 0.1334 & 0.146 & 0.147 & 0.08691 & 0.0955 & 0.096 & 0.07475 & 0.0821 & 0.082 & 0.06921 & 0.0760 & 0.077 \\ \hline
        1 & 0.07654 & 0.0776 & 0.0777 & 0.04965 & 0.0503 & 0.0504 & 0.04266 & 0.0433 & 0.0434 & 0.03949 & 0.0400 & 0.0406 \\ \hline
        2 & 0.04672 & 0.0468 & 0.0466 &  0.03017 & 0.0302 & 0.0302 & 0.02589 & 0.0259 & 0.0259 & 0.02396 & 0.02400 & 0.0241 \\ \hline
 $\infty$ & 0.02329 & 0.02329 & 0.02329 & 0.01496 & 0.01496 & 0.01496 & 0.01282 & 0.01282 & 0.01282 & 0.01185 & 0.01185 & 0.01189 \\ \hline
       -2 & 0.01043 & 0.01043 & 0.0106 & 0.006669 & 0.006669 & 0.00674 & 0.005710 & 0.005710 & 0.00571 & 0.005278 & 0.005278 & 0.00527 \\ \hline
       -1 & 0.005022 & 0.005022 & 0.0049 & 0.003204 & 0.003204 & 0.0032 & 0.002742 & 0.002742 & 0.0026 & 0.002534 & 0.002534 & 0.0023 \\ \hline
     -1/2 & 0.001703 & 0.001703 & 0.002 & 0.001085 & 0.001085 & 0.001 & 0.0009282 & 0.0009282 & 0.001 & 0.0008579 & 0.0008579 & 0.001 
  \end{tabular}       
  \caption{{\bf Fermion:} Finite angle results for $a_\alpha(\theta)$ for the Dirac fermion at \ren index $\alpha=1,2,3,4$. 
The series and ansatz \req{new-ansatz} results
are obtained using field theory, whereas the lattice results are obtained numerically using the direct method.
The truncated series includes terms up to $(\theta-\pi)^{14}$, and can be taken as a lower bound for $a_\alpha(\theta)$.} 
  \label{tab:fermion}    
\end{table*} 

\begin{figure}
 \includegraphics[width=\linewidth]{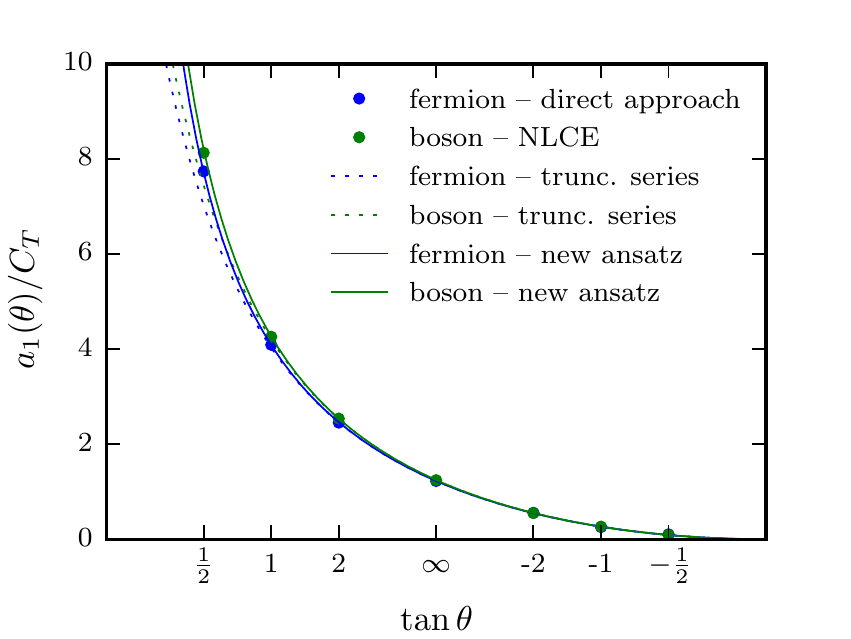}
 \caption{Angle dependence for the von Neumann ($\alpha=1$) entropy. 
The lattice data agree very well with our new field theory based ansatz, \req{new-ansatz}. 
For the field theory ansatz and series curves, see Section~\ref{sec:QFT}.}   
\label{fig:anglesplot} 
\end{figure}

Consider first the corner coefficient in the von Neumann entropy ($\alpha\!=\!1$) plotted in Fig.~\ref{fig:anglesplot}. 
By Eq.~\eqref{smooth-CT}, $a_1(\theta)$ is proportional to the central charge $C_T$ of the stress-tensor of the CFT\cite{Bueno:2015,Faulkner:2015aa} in the smooth limit.
We therefore plot the ratio $a_1(\theta)/C_T$, where\cite{Osborn:1994} $C_T = {3}/{16 \pi^2}$ .
The numerically obtained data points are shown in solid blue and green. 
In addition, we plot the series expansion about the smooth limit obtained using the methods of Ref.~\onlinecite{Casini:2009} 
up to order $14$ (fermion) and $16$ (boson) in $(\theta-\pi)$. 
Our numerical lattice data is in good agreement with the truncated series for small $(\pi-\theta)$. 
At $\theta\! =\! \arctan(1/2)\approx 0.46$, however, there is a $6\%$ (boson) and $9\%$ (fermion) difference between the lattice data and the series.
In contrast, our new field theory ansatz \req{new-ansatz} only deviates by $1\%$ from the lattice data at this angle for both theories.  See 
Tables~\ref{tab:boson}-\ref{tab:fermion} for the comparison at all angles.    

\begin{figure*}[t]
    \centering
    \includegraphics[width=0.49\linewidth]{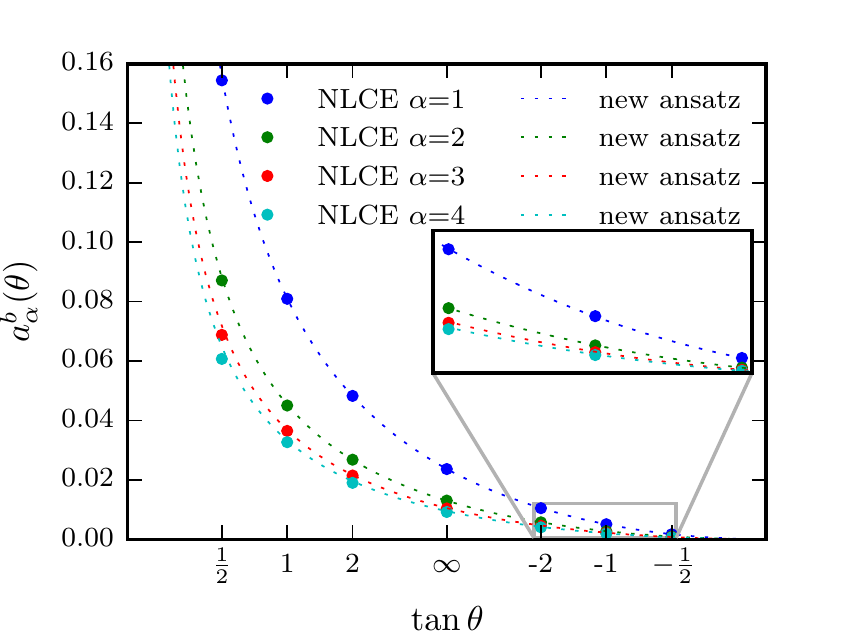}
    \includegraphics[width=0.49\linewidth]{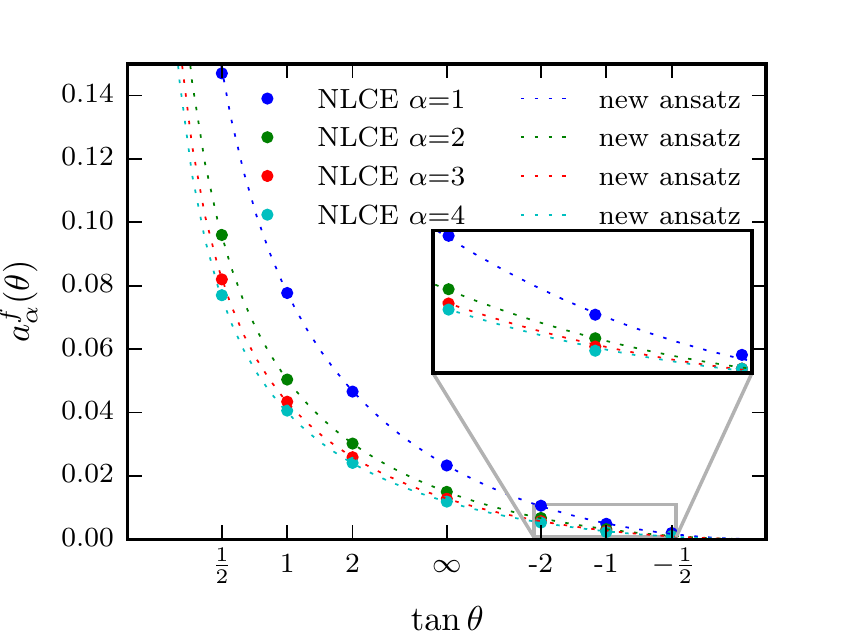}
    \caption{Corner entanglement for \Renyi entropies for the free boson (left panel) and the Dirac fermion (right panel). The numerical data is compared to our new field theory ansatz, \req{new-ansatz}, at $M\!=8$ (7) for the boson (fermion).   
 }
    \label{fig:angles234}
\end{figure*}

We next examine the corner coefficient for the \ren entropies with arbitrary $\alpha$.
Limited numerical results are available for comparison from the literature. For the boson: 
$a_1^b(\pi/2)\!=\! 0.02366$,\cite{Casini:2007,Casini:2008,Laflorencie:2015}
$a_2^b(\pi/2)\!=\! 0.0128(2)$,\cite{Casini:2007,Laflorencie:2015} 
$a_3^b(\pi/2)\!=\! 0.0100(2)$,\cite{Casini:2007,Laflorencie:2015}, 
$a_4^b(\pi/2)\!=\! 0.0086(2)$.\cite{Laflorencie:2015} 
In the first 3 cases the agreement with our findings (see Table~\ref{tab:boson}) is excellent, while it is less so for the last one.
There we expect that our new field theory answer is the most accurate: $0.00946$.
For the Dirac fermion, $a_1(\pi/2)=0.02329$,\cite{Casini:2008} again in agreement with our finding, Table~\ref{tab:fermion}. 
In the less explored regime of $\alpha\!<\! 1$,
Ref.~\onlinecite{Nobili:2016} indirectly obtains $a_{1/2}^b(\pi /2 ) = 0.058(2)$ by using
the entanglement \emph{negativity}.  For this value of $\alpha$, our method yields $0.058$, in agreement.
Further calculations are summarized in  
Fig.~\ref{fig:angles234}, where we compare $a_\alpha(\theta)$ for $\alpha=1,2,3,4$, for both  
bosons and fermions.  For the \ren entropies, a simple lower bound has been derived in Ref.~\onlinecite{Bueno:2016},  
which we find is strictly obeyed in all cases.   

\begin{figure*}[b]
    \centering
    \includegraphics{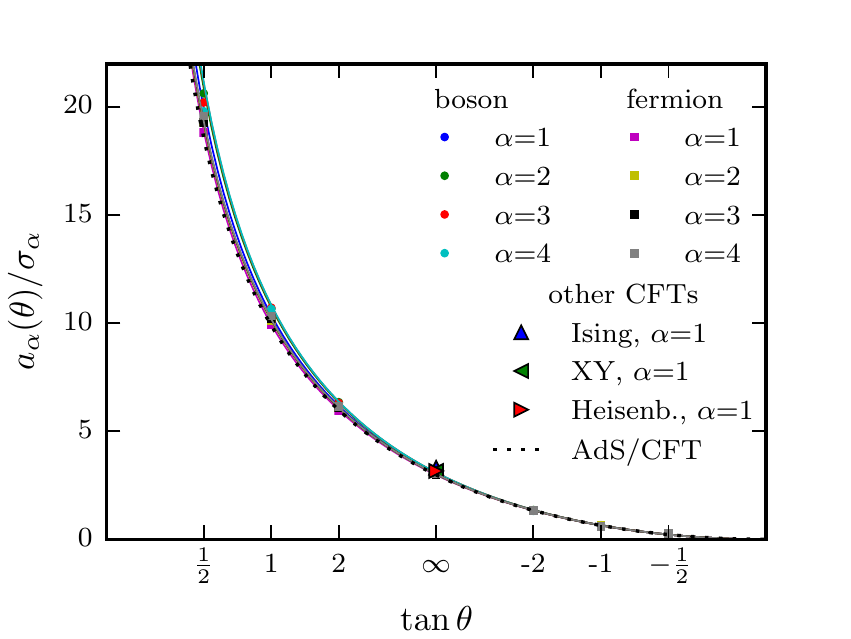}  
    \includegraphics{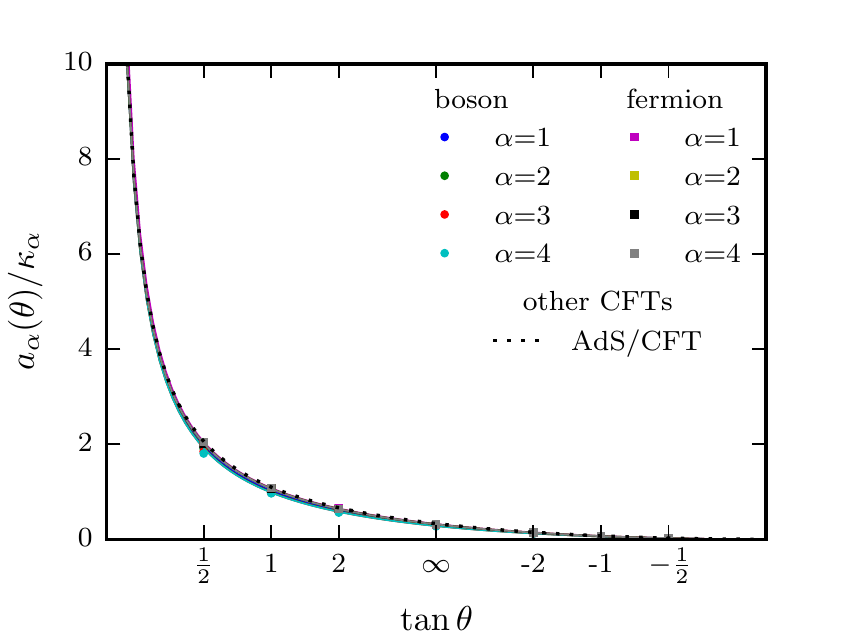}  
    \caption{
Collapsing the corner entanglement after dividing $a_\alpha(\theta)$ by the smooth limit coefficient $\sigma_\alpha$ (left panel). 
The solid lines are obtained using field theory \req{new-ansatz}, while the markers are lattice data.
We also show the $\pi/2$ data for the $O(N)$ Wilson-Fisher QCPs with Ising, XY and Heisenberg symmetry. 
The AdS/CFT result corresponds to a strongly-interacting CFT.\cite{Hirata:2007} The solid lines show the new ansatz. 
Another way to collapse the data is to divide $a_\alpha(\theta)$ by the sharp limit coefficient $\kappa_\alpha$ (right panel), which achieves a better
agreement for $\theta < \pi / 2$.
}
    \label{fig:masterplot_both}
\end{figure*}

The first smooth limit expansion coefficients $\sigma_\alpha$ in Eq.~\eqref{smooth1} have been computed in 
Refs.~\onlinecite{Elvang:2015,Bueno:2015JHEP} 
for the free boson and the Dirac fermion for a few \Renyi indices.
Plotting our numerical data for $a_\alpha(\theta)$ divided by the corresponding $\sigma_\alpha$ in Fig.~\ref{fig:masterplot_both}a, 
we see an almost complete collapse of the different \Renyi EE onto a single curve.
This surprising result suggests that $\sigma_\alpha$ essentially determines the values of $a_\alpha(\theta)$,
even at angles near $\pi/2$. At smaller angles, the spread among the different curves 
becomes clear. In that regime 
it is more sensible to examine $a_\alpha(\theta)/\kappa_\alpha$, where $\kappa_\alpha$ 
determines the $1/\theta$ divergence, as is shown in Fig.~\ref{fig:masterplot_both}b. 

In addition, our data indicates that $a_1^b(\theta)\! >\! a_1^f(\theta)$, while the inequality is reversed
for $\alpha=2,3,4$: $a_\alpha^b<a_\alpha^f$. 
In general, our results show that the boson corner term, when normalized by the smooth limit coefficient, 
exceeds that of the fermion for $\alpha=1,2,3,4$: 
\begin{align}
  a_\alpha^b(\theta)/\sigma_\alpha^b > a_\alpha^f(\theta)/\sigma_\alpha^f\,.
\end{align}
In Ref.~\onlinecite{Bueno:2015}, it was noted that the Dirac fermion and a family of strongly-coupled critical theories
described by the holographic correspondence, and other models, have $a_1(\theta)/\sigma_1$ less than that of the free boson.
It is possible that this applies to all CFTs, but it remains to be shown. Here we raise the stronger question:
is $a_\alpha(\theta)/ \sigma_\alpha$ an upper bound for all CFTs, at least in a range of $\alpha$? It will be interesting
to resolve this question, which is related to the physical meaning of $\sigma_\alpha$ for $\alpha\neq 1$.
For the free boson and Dirac fermion $\sigma_\alpha$ is related to the scaling dimension of 
twist operators,\cite{Bueno:2015JHEP,Dowker:2015,Dowker:2015b} and it is not clear whether this holds in general. 

All of our numerical data points in the figures are summarized in the \emph{lattice} columns of Tabs. \ref{tab:boson} and \ref{tab:fermion}.     
Their strict confidence intervals are hard to rigorously determine, as there are systematic errors related to the restricted size of the lattices and the various options available in the 
data fitting procedures.
In the tables we show the digits that we found to be robust against variations in the fitting ranges.
The accuracy varies from two digits for smooth angles at $\alpha = 3,4$ to four digits for $\theta = \pi / 2$.
We estimate that the third significant digit of the resulting corner coefficient contains the systematic uncertainty.

\subsubsection{Boson-fermion duality} \label{BF}      

An exact duality between the smooth-limit corner coefficient $\sigma_\alpha$ of the boson and fermion 
was identified in Ref.~\onlinecite{Bueno:2015}:  
\begin{equation}
    \alpha^2 \sigma^{f}_\alpha = \sigma^{b}_{1/ \alpha}.
    \label{eq:duality}
\end{equation}
In this section, we study the extent to which the duality holds at angles $\theta < \pi$.
That is, we ask whether 
  \begin{align} \label{strong-duality}
    \alpha^2 a_\alpha^f(\theta) \stackrel{\text{?}}{=} a_{1/\alpha}^b(\theta)
  \end{align}
holds also at generic $\theta$. 

\begin{figure*}[h]
    \includegraphics[width=0.49\linewidth]{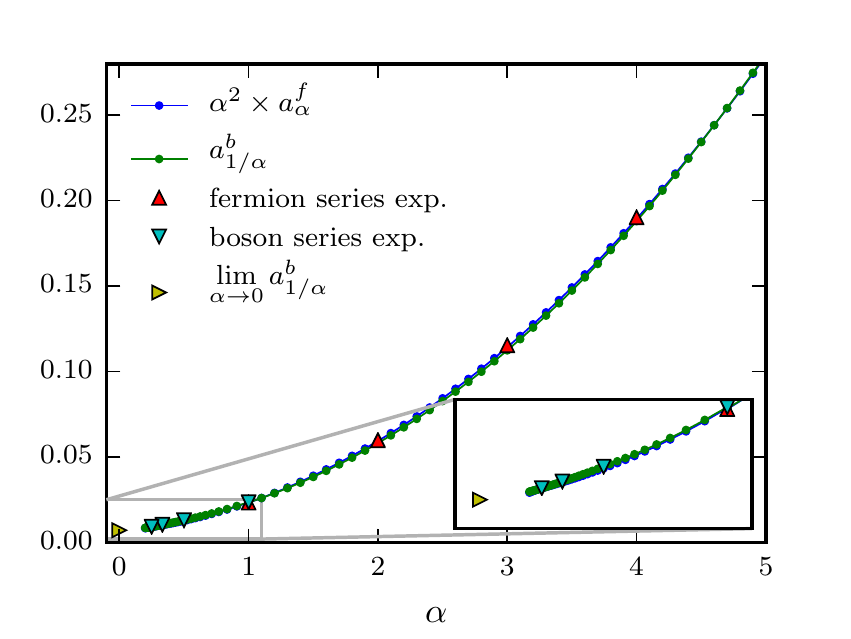}
    \includegraphics[width=0.49\linewidth]{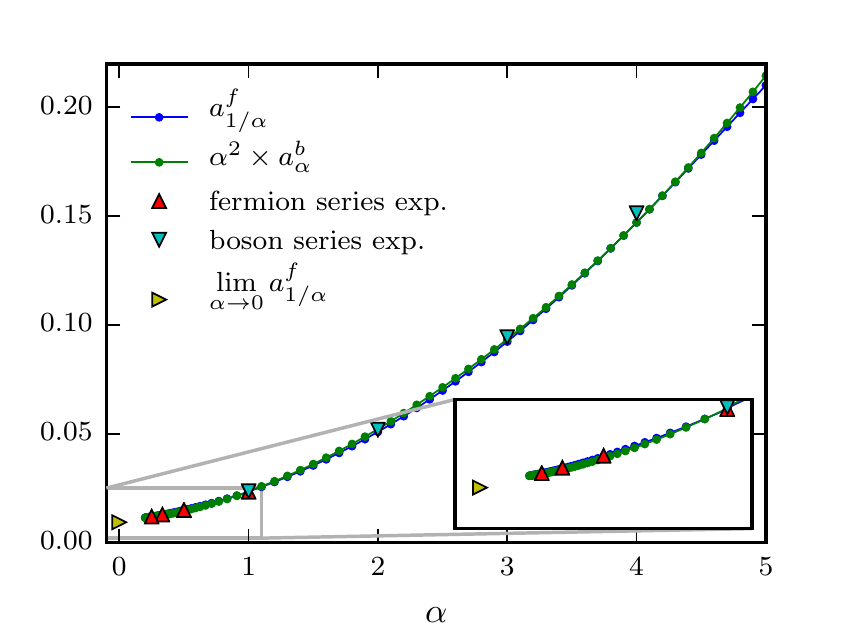}
    \caption{Testing the boson-fermion duality at $\theta\!=\!\pi /2$.
The data for fermions has been obtained by a fit to the area law using an exact diagonalization of the subsystem correlation matrix. For bosons, the NLCE was used. 
$a_\infty(\pi/2)$ was computed using the BMW ansatz\cite{Bueno:2015JHEP} to be $0.0093\, (0.00715)$ for the fermion (boson) respectively. 
Both plots display the same data 
    with different $\alpha$ prefactors, in order to resolve clearly the duality for both sides of $\alpha=1$. }
    \label{fig:duality} 
\end{figure*}

To wit, we plot the two sides of Eq.~\eqref{strong-duality} for \Renyi indices $\alpha \in [\frac15 \dots 5]$ at $\theta=\pi/2$ in Fig.~\ref{fig:duality}.
In the left panel, we test Eq.~\eqref{strong-duality}, while in the right panel, we test the same equation when $\alpha \to 1/\alpha$. 
The plots show that the duality approximately holds even at $\theta=\pi/2$.
Moreover, the data are in good agreement with exact results available in the literature for some integer-valued \Renyi entropies (see Sec.~\ref{Sec:AngleDependenceCorner} for details). 
We note that at $\alpha=1$, we have\cite{Bueno:2015,Elvang:2015}
\begin{align} 
  \sigma_1^f = \sigma_1^b = \frac{1}{128}
\end{align}
  
 \onecolumngrid

\newpage

\twocolumngrid
\noindent
If the duality were to hold for all $\theta$ at $\alpha=1$, $a_1(\theta)/C_T$ would be the same for the boson and the fermion. 
Fig.~\ref{fig:anglesplot} plots this function for the two theories.  
The results can hardly be distinguished on the interval $[\pi/2,\pi]$, showing that the duality is robust in this range.
However, the growing discrepancy between the two curves as $\theta \rightarrow 0$ proves that the duality breaks down in that limit. Indeed, the coefficients of the $1/\theta$ divergence, \req{sharp}, clearly differ: $\kappa_1^b=0.0794$
 and $\kappa_1^f\! =\! 0.0722$.\cite{Casini:2009}    

The precision of the computation is expected to be better for $\alpha \geq1$ as compared to $\alpha<1$. 
Computing the values of the \Renyi entropies involves a sum over many eigenvalues 
raised to the power of the \Renyi index.
Per default, the numerical precision of floating point numbers and operations on them is 16 digits in the mantissa.
However, if we compute, for example, the $\frac14$-\Renyi entropy the precision is reduced to four~digits 
after taking the $\frac14$th power in the sum in Eq.~(\ref{eq:fermionentropy}). 
These values enter the outer sum of the entropy formula and are then further relied upon 
in the NLCE or the finite-size fit to the area law in the direct approach.
As a result, the precision of the corner term for small \Renyi indices is reduced such that, for $\alpha < 0.5$, it is 
difficult to find asymptotic behavior for both fermions and bosons using standard precision floating point numbers.

We therefore employ arbitrary precision numerics to obtain the \Renyi entropies for all $\alpha<1$. 
This leads to a dramatic reduction on the achievable system size (or cluster order) as these routines
--- in particular the diagonalizations --- are significantly slower. 
Since this problem does not apply for $\alpha \geq 1$ we combine the corner term curves for both
the fermions and the boson from two different calculations --- high precision but lower order for $\alpha<1$ 
and standard precision but higher order for $\alpha \geq 1$.  
This likely explains the slight deviation from the duality which becomes apparent near $\alpha=5$ ($\alpha=1/5$ resp.).

We can further test the duality at $\theta\!=\!\pi/4$ by using the boson results 
of Ref.~\onlinecite{Nobili:2016}: $a_{1/2}^b(\pi /4 ) = 0.0195(6)$ and $a_{1/2}^b(3\pi /4 ) = 0.012(1)$. 
We can compare these with the ``dual'' fermion results: $2^2 a^f_{2}(\pi /4) = 0.0202(4) $ and $2^2 a^f_{2}(3\pi /4) = 0.0128(4)$.  
Both sets of values agree within the estimated numerical uncertainty, which is contained in the last given digit in Tab.~\ref{tab:fermion}.
This further supports the statement that the duality works well beyond the smooth limit.   

Finally, we are unable to quantitatively determine how large the discrepancy between the two sides of Eq.~\eqref{strong-duality} is. 
The finite system size and finite floating point precision contribute to the error in the numerical computation.
As the errors likely behave differently at different opening angles and for the two different field theories, an extrapolation to the thermodynamic limit is tricky.
Our results nevertheless show that the duality is robust, in particular for the natural lattice angle of $\theta = \pi / 2$.

\section{Discussion} \label{sec:discussion}
We have presented a comprehensive analytical and numerical study of the universal term in the $\alpha$-\ren
entropies that arise due to the presence of a corner in the entangling boundary.
We have focused on two non-interacting quantum critical theories in (2+1) space-time dimensions: 
the free relativistic boson and Dirac fermion. 
This corner contribution arises as a logarithmic correction to the area law, with coefficient 
$a_{\alpha}(\theta)$ being the universal quantity of interest. 

First, we used a field theory approach\cite{Casini:2007,Casini:2009} based on an expansion around the smooth corner limit. 
Using this expansion we obtained a new set of strong lower 
bounds for $a_{\alpha}(\theta)$. 
We also constructed a simple closed-form ansatz, exact in the limit of smooth and sharp corners, 
that gives high-precision estimates for the corner coefficient at all angles.  

We then employed a lattice numerical approach to obtain $a_{\alpha}(\theta)$ for angles amenable to discretization on a square 
lattice, away from the smooth corner limit.  For free theories,
the reduced density matrix over region $A$ can be obtained directly from
the lattice correlation functions of the fields and their conjugate momenta.
We use this as the basis for a numerical diagonalization scheme which can  
extract $a_{\alpha}(\theta)$ in the thermodynamic limit through two finite-size scaling methods.  
One is a direct calculation of the $\alpha$-\ren entropies for a closed polygonal region $A$ embedded in a finite-size torus, which we find gives the most accurate results for the fermion.  The other is a numerical linked-cluster expansion (NLCE), which scales 
most favorably for the boson.

Our combined continuum and lattice approach allows us to make a wide range of conclusions 
about $a_\alpha(\theta)$. First, we find that the coefficient $\sigma_\alpha$ that determines
the smooth limit behavior, $a_\alpha(\theta\approx\pi)=\sigma_\alpha\,(\theta-\pi)^2$, essentially
controls $a_\alpha(\theta)$ at angles as small as $\pi/2$. This is illustrated by the near collapse of $a_\alpha(\theta)/\sigma_\alpha$, 
Fig.~\ref{fig:masterplot_both}. 
This also explains the robustness of the boson-fermion duality beyond the smooth limit, where it was originally
derived.\cite{Bueno:2015JHEP} 
On the other hand, it is known that $\sigma_1$ is a measure of the gapless degrees of freedom of the system
by virtue of giving the stress tensor central charge, \req{smooth-CT}.
Corner entanglement thus offers a practical way to probe gapless modes in interacting lattice models, 
where a quasiparticle description often does not exist. 
This could be of immediate use in 
lattice models that contain quantum phase transitions with unknown low-energy descriptions, such as deconfined quantum critical
points.\cite{Senthil:2004,Sandvik:2010d,Nahum_2015}

It would also be of interest to apply the numerical methods used here to 
other theories.  Quantum critical systems with $z\neq 1$, 
such as the quadratic band touching model of fermions, would be a simple example. 
Further, our methods and ideas can easily be generalized to three\cite{Devakul_2014}
(and higher) spatial dimensions, where very little is known. 
We expect that numerical studies of
\ren entropies in quantum critical systems
will reveal rich universal behavior centering around local corners and vertices in the entangling boundary.

\begin{acknowledgments}   
We are extremely grateful to H.~Casini for sharing some of his unpublished field theory results
and for numerous discussions. 
We also acknowledge crucial discussions with P.~Bueno, P.~Fendley, and R.~Myers. 
J.H. thanks the Perimeter Institute for Theoretical Physics for hospitality 
and acknowledges support from the Bonn-Cologne Graduate School of Physics and Astronomy. 
Funding for J.H. was provided by the Institutional Strategy of the University of Cologne within the German Excellence Initiative.
L.H.S. is partially funded by the Ontario Graduate Scholarship. W.W.K.\/ is partially funded by a
fellowship from NSERC.  
R.G.M.~acknowledges support from NSERC, the Canada Research Chair program, and the Perimeter Institute. 
The simulations were performed on the computing facilities of SHARCNET and on the Perimeter Institute HPC. 
 Research at the Perimeter Institute is supported by the Government of Canada through Industry Canada and by the Province of Ontario through the Ministry of Economic Development \& Innovation.
\end{acknowledgments}

%
%
\onecolumngrid
\appendix
\section{Smooth limit coefficients} \label{app:smooth_coeffs}
In this appendix, we give the smooth limit coefficient $\sigma_\alpha^{(p)}$ for both the boson and Dirac fermion,
at $\alpha=1,2,3,4$, see Tables~\ref{tab:scalar-CFT}-\ref{tab:fermion-CFT}, respectively.  
These were evaluated to high precision using the methods of Ref.~\onlinecite{Casini:2009}.    
The closed-form results for the quartic coefficient $\sigma_\alpha'$ are new. An outline of the calculation 
of $\sigma_1'$ for the boson is given in Appendix~\ref{app:sigma1}.

It is worth emphasizing that the computation is substantially easier for $\alpha\!>\! 1$ compared with 
the von Neumann $\alpha\!=\! 1$ case, since the former requires one less integral compared with the latter, \req{corner-Ren-cs} versus \req{corner-EE-cs}.   
In practice this means that one can readily obtain $50$ digits of accuracy for $\alpha>1$. In Tables~\ref{tab:scalar-CFT}-\ref{tab:fermion-CFT} we have only quoted 
a few digits; the last one was rounded.     

Figure~\ref{fig:sig_b} shows the coefficients for the boson normalized by their asymptotic value.
We note that all the coefficients decrease with increasing \ren index in the range studied.     
\begin{table*}[h]
  \centering
  \begin{tabular}{c||c|c|c|c|c|c|c|c}     
   $\alpha$ & $\sigma_\alpha$ & $\sigma_\alpha'$ & $\sigma_\alpha''\times 10^5$ & $\sigma_\alpha^{(3)}\times 10^6$ & $\sigma_\alpha^{(4)}\times 10^7$ & $\sigma_\alpha^{(5)}\times 10^8$ 
& $\sigma_\alpha^{(6)}\times 10^9$ & $\sigma_\alpha^{(7)}\times 10^{10}$ \\ 
    \hline \hline
    1 & $\frac{1}{128}$  & $\frac{20+ 3\pi^2 }{9216\pi^2}$ & $5.34655497$ & $5.40160621$ & $5.45758486$
& $5.51156763$ & $5.57181927$ & $5.63580458$ \Ts\Bs \\ \hline 
    2 & $\frac{1}{24\pi^2}$ & $\frac{5+\pi^{2}}{480\pi^4}$ & $3.11534753$ & $3.12412616$ & $3.14738400$ & $3.17722233$ & $3.21122771$ & $3.24805958$ \Ts\Bs \\ \hline
    3 & $\frac{1}{54\sqrt{3} \pi}$ & $\frac{70 \sqrt 3 \pi-81}{116640 \pi^2}$ & $2.55467090$ & $2.560 911 69$ & $2.579 241 60$ & $2.603 271 34$ 
& $2.630 863 69$ & $2.660 842 50$ \Ts\Bs \\ \hline
    4 & $\frac{8 + 3 \pi}{576 \pi^2}$ & $2.356 888 62\!\cdot\! 10^{-4}$ & $2.312 613 23$ & $2.318 443 02$ & $2.335 037 67$ & $2.356 761 91$ 
& $2.381 707 82$ & $2.408 816 42$ \Ts\Bs     
  \end{tabular}   
  \caption{{\bf Boson:} First 8 smooth limit coefficients for the boson at \ren indices $\alpha=1,2,3,4$. All shown digits are significant.
The closed-form answers for $\sigma_\alpha'$ are new.}   
  \label{tab:scalar-CFT}    
\end{table*}
  
\begin{table*}[h]
  \centering
  \begin{tabular}{c||c|c|c|c|c|c|c}    
   $\alpha$ & $\sigma_\alpha$ & $\sigma_\alpha'$ & $\sigma_\alpha''\times 10^5$ & $\sigma_\alpha^{(3)}\times 10^6$ & $\sigma_\alpha^{(4)}\times 10^7$ & $\sigma_\alpha^{(5)}\times 10^8$ 
& $\sigma_\alpha^{(6)}\times 10^9$ \\
    \hline \hline
    1 & $\frac{1}{128}$  & $\frac{16 + 3\pi^2}{9216\pi^2}$ & $4.812 997 0$ & $4.855 231 7$ & $4.917 335 3$ & 
$4.977 709 7$ & $5.041 144 7$ \Ts\Bs \\ \hline 
    2 & $\frac{1}{64\pi}$  & $\frac{35 \pi - 8}{30720 \pi^2}$ & $3.194 260 62$ & $3.186 737 87$ & $3.215 499 55$ & $3.254 310 08$ 
& $3.296 349 26$ \Ts\Bs \\ \hline 
    3 & $\frac{5}{216 \sqrt 3 \pi}$  & $\frac{410 \sqrt 3 \pi -891}{466560 \pi^2}$ & $2.758 583 23$ & $2.747 062 75$ 
& $2.769 403 30$ & $2.801 826 52$ & $2.837 668 69$ \Ts\Bs \\ \hline 
    4 & $\frac{1 + 6 \sqrt 2}{768 \pi}$  & $2.700 522 31\!\cdot\! 10^{-4}$  & $2.558 313 57$ & $2.546 096 28$ & $2.565 979 99$ 
    & $2.595 634 62$ & $2.628 674 63$ \Ts\Bs  
  \end{tabular}    
  \caption{{\bf Fermion:} First 7 smooth limit coefficients for the massless Dirac fermion for \ren indices $\alpha=1,2,3,4$. All shown digits are significant. 
The closed-form answers for $\sigma_\alpha'$ are new.}   
  \label{tab:fermion-CFT}    
\end{table*}

\section{Exact calculation of $\sigma_\alpha'$}
\label{app:sigma1}
Here we show the key steps in the exact calculation of the coefficient $\sigma_1'$ of $(\theta-\pi)^4$ for the
boson.  It is defined via the double integral
\begin{align}
  \sigma_1' = \int_0^\infty dt \int_{1/2}^\infty dm \, f(t,m) ,
\end{align}
where the integrand is a fairly complicated function that can be obtained from the system of equations
for $H_a$ derived by Casini and Huerta.\cite{Casini:2009} Following similar steps that Elvang and Hadjiantonis
took in their derivation\cite{Elvang:2015} of the leading coefficient $\sigma_1$ we make the change of variables
\begin{align}
  u = \sqrt{m^2 -\tfrac{1}{4}}.
\end{align}
After numerous simplifications, we obtain 
\begin{multline}
  f(t,u) = \frac{u^2 \text{sech}^2(\pi  t) \sinh (\pi  (t+u)) }{48 (\cosh (2 \pi  t)-\cosh (2 \pi  u))^3} \\
  \times \Bigg[ (\cosh (2 \pi  t)-\cosh (2 \pi  u))^2 \left(3 \left(t^2-u^2\right)+\left(3 t^2-3 u^2-1\right) \cosh (2 \pi  u)+\cosh (2 \pi  t)\right) \text{csch}(\pi  (t+u)) \\
  -\frac{2 \cosh (\pi  u)}{u} \sinh (\pi  (t-u)) \left(\left(-3 t^2+3 u^2-1\right) \cosh (2 \pi  t)+\left(-3 t^2+3 u^2+1\right) \cosh (2 \pi  u)-6 t^2+6 u^2\right) \\
  \times \left(2 \pi  \left(u^2-t^2\right) \sinh (\pi  u)+\cosh (\pi  u) \left(\pi  \left(t^2-u^2\right) (\coth (\pi  (t+u))-\coth (\pi  (t-u)))
      +2 u\right)\right) \Bigg].
\end{multline}
We further find it convenient to change variables to the center-of-mass and relative coordinates
\begin{align}
  R = (t+u)/2\,, \quad r = u-t\,.
\end{align} 
This substitution leads to certain simplifications; for instance, $\cosh(2\pi u)-\cosh(2\pi t)=2 \sinh(\pi r) \sinh( 2\pi R)$
factorizes.
We then separate the integrand into more manageable parts and integrate term-wise. 
However, care must be used
since certain terms lead to divergent integrals when considered by themselves.
In the process, one encounters integrals of the type
\begin{align}
  \int_0^\infty dR \int_{-\infty}^{\infty} dr \frac{r^2 R^2}{\sinh^2(\pi r)\sinh^2(2\pi R)} = \frac{1}{16}\times \frac{1}{(3\pi)^2},
\end{align}
which factorize.  
The final answer is  
\begin{align}
  \sigma_1'= \frac{20+3\pi^2}{9216\pi^2}.
\end{align}

\section{Bounds in strongly coupled holographic theories} \label{app:ads} 
\begin{figure}  
  \center 
  \includegraphics[scale=.415]{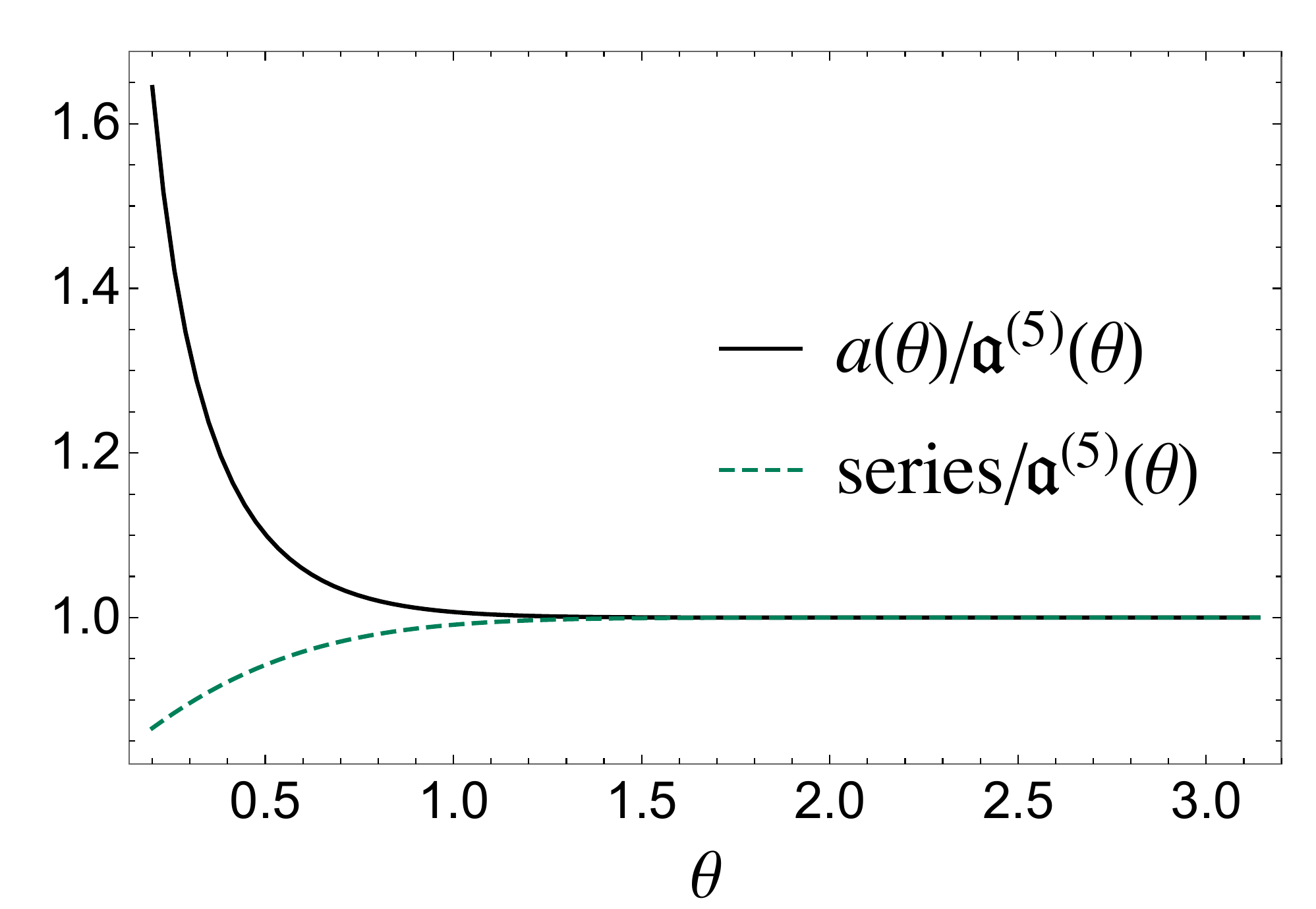}   
  \caption{Solid: corner function for holographic theories,\cite{Hirata:2007} normalized by the lower bound $\mathfrak a^{(5)}$.
Dashed: the series including terms up to $(\theta-\pi)^{10}$, normalized by the lower bound.} 
  \label{fig:ads}  
  \centering    
\end{figure}     
 
We test the new lower bound $\mathfrak a_1^{(M)}(\theta)$ using the gauge/gravity (AdS/CFT) duality.
We shall drop the subscript $\alpha=1$ for the remainder of this appendix.
A beautiful prescription\cite{RT} allows one to compute the corner function $a(\theta)$ for a family of quantum
critical theories,\cite{Hirata:2007} which are (supersymmetric) CFTs. We examine the bound for $M\!=\!5$, $\mathfrak a^{(5)}(\theta)$. 
We can use the
first 5 smooth-limit coefficients computed analytically in Ref.~\onlinecite{Bueno:2016}. (The \emph{ratios} of the first
4 coefficients are numerically given in Ref.~\onlinecite{Casini:2008}, and agree with the exact answers.\cite{Bueno:2016})    
The result is plotted in Fig.~\ref{fig:ads}, where we verify that the hierarchy \req{hierarchy} holds: 
\begin{align} \label{hierarchy2}
  \sum_{p=1}^{5}\sigma_\alpha^{(p-1)}\,(\theta-\pi)^{2p} \, < \, \mathfrak a^{(5)}(\theta) \, < \,  a(\theta)\,.
\end{align}   
The bound and series are very accurate up to $\theta\approx \pi/2$.

%
%

\bibliography{Biblo}   

\end{document}